\documentclass[a4paper,12pt]{article}

\usepackage{amsmath,amssymb,amsthm,mathrsfs}
\usepackage{latexsym}
\usepackage{cite}
\usepackage{tikz}
\usetikzlibrary{matrix,arrows}
\tikzset{node distance=3cm, auto}
\numberwithin{equation}{section}

\newcommand{\del}{\partial}

\newcommand{\Dcal}{\mathcal{D}}
\newcommand{\e}{\epsilon}
\newcommand{\lambdabar}{\bar{\lambda}}
\newcommand{\sigmabar}{\bar{\sigma}}
\newcommand{\xibar}{\bar{\xi}}

\newcommand{\Tr}{\mathrm{Tr}}
\newcommand{\Zbar}{\bar{Z}}
\newcommand{\zbar}{\bar{z}}
\newcommand{\qbar}{\bar{q}}
\newcommand{\Fbar}{\bar{F}}
\newcommand{\psibar}{\bar{\psi}}
\newcommand{\Ybar}{\bar{Y}}
\newcommand{\chibar}{\bar{\chi}}
\newcommand{\mubar}{\bar{\mu}}

\newcommand{\Wbar}{\bar{W}}
\newcommand{\Qbar}{\bar{Q}}
\newcommand{\zetaa}{\zeta_1^{(1)}}
\newcommand{\zetab}{\zeta_1^{(2)}}
\newcommand{\zetac}{\zeta_2^{(1)}}
\newcommand{\zetad}{\zeta_2^{(2)}}
\newcommand{\Phibar}{\bar{\Phi}}
\newcommand{\Fcal}{\mathcal{F}}
\newcommand{\Bcal}{\mathcal{B}}
\newcommand{\Ncal}{\mathcal{N}}
\newcommand{\Tcal}{\mathcal{T}}
\newcommand{\et}{\tilde{\e}}
\newcommand{\R}{\mathbb{R}}
\newcommand{\C}{\mathbb{C}}
\newcommand{\Z}{\mathbb{Z}}
\newcommand{\Zcal}{\mathcal{Z}}
\newcommand{\Pcal}{\mathcal{E}}

\newcommand{\cbar}{\bar{c}}

\newcommand{\be}{\begin{equation}}
\newcommand{\ee}{\end{equation}}
\newcommand*\circled[1]{\tikz[baseline=(char.base)]{
            \node[shape=circle,draw,inner sep=2pt] (char) {#1};}}

\begin{document}
\renewcommand{\thefootnote}{\fnsymbol{footnote}}
\begin{titlepage}

\vspace*{10mm}

\begin{center}
{\Large \textbf{
Exact Results in Quiver Quantum Mechanics\\[0.8em]
and\\[1.0em]
BPS Bound State Counting
}}
\vspace*{12mm}

\normalsize{Kazutoshi Ohta\footnote{E-mail: kohta@law.meijigakuin.ac.jp} and Yuya Sasai\footnote{E-mail: sasai@law.meijigakuin.ac.jp}}

\vspace*{12mm}
\textit{
Institute of Physics, Meiji Gakuin University, Yokohama 244-8539, Japan  
}

\end{center}

\vspace*{15mm}

\begin{abstract}

We exactly evaluate the partition function (index) of ${\cal N}=4$ supersymmetric quiver quantum mechanics
in the Higgs phase
 by using the localization techniques.
We show that the path integral is localized at the fixed points, which are obtained by
solving the  BRST equations, and D-term and F-term conditions.
We turn on  background gauge fields of R-symmetries
for the chiral multiplets corresponding to the arrows between quiver nodes,
but the partition function does not depend on these R-charges.
We give explicit examples of the quiver theory including
a non-coprime dimension vector.
The partition functions completely agree with the mathematical formul{\ae} of the Poincar\'{e} polynomials
($\chi_y$-genus)
and the wall crossing for the quiver moduli spaces .
We also discuss exact computation of the expectation values of supersymmetric ($Q$-closed) Wilson loops
in the quiver theory.

\end{abstract}

\end{titlepage}

\newpage
\renewcommand{\thefootnote}{\arabic{footnote}}
\setcounter{footnote}{0}

\section{Introduction}

A count of the number of the BPS bound states (index) in supersymmetric theory,
including supergravity and superstring theory, is important problem
to understand quantum properties of the space-time.
For example, the counting of the BPS bound states on D-branes
makes relation to the Bekenstein-Hawking entropy of a black hole \cite{Strominger:1996sh}.
Denef proposed that  the BPS bound states (multi-centered black holes) in Calabi-Yau compactifications of
string theory is interpreted by a suitable quiver quantum mechanics  and discussed the correspondence between the bound states of the wrapped D-branes around the Calabi-Yau manifold (Higgs picture)  and the multi-centered  bound states in the supergravity (Coulomb picture) \cite{Denef:2002ru}.
These correspondences have been checked for many cases so far and imply that string theory gives the proper description of quantum gravity. 

The quantity of the number of the
BPS bound states (index) is also interesting mathematically, since it relates to
the topological invariants of the manifolds.
Therefore, a lot of interesting formul{\ae}, which compute the number of the BPS bound states or
topological invariants, have been derived. In mathematics, topological properties of
the quiver moduli spaces are investigated in \cite{Reineke,Weist} for example,
and they come to fruition of the so-called ``wall crossing formula'' by Joyce and Song
\cite{Joyce:2008pc}, and Kontsevich and Soibelman \cite{Kontsevich:2008fj}.
The wall crossing formula is physically interpreted and rederived 
in the Coulomb branch of the quiver quantum mechanics \cite{deBoer:2008zn, Manschot:2010qz}.
There are many important developments around the wall crossing formula
\cite{Gaiotto:2008cd,Nishinaka:2010qk,Manschot:2011xc,Lee:2011ph,Nishinaka:2011nn,Kim:2011sc,Sen:2011aa,Bena:2012hf,Lee:2012sc,Lee:2012naa,Manschot:2012rx,Lee:2013yka,Manschot:2013sya,Manschot:2014fua}. (See also the review of the developments \cite{Pioline:2013wta}.)

In this paper, we would like to evaluate the partition function of the quiver quantum mechanics
in the Higgs phase   by using the localization method.
Recently, the index in two-dimensional supersymmetric gauge theory is computed by
the localization and Jeffrey-Kirwan (JK) residue operation \cite{Benini:2013nda,Benini:2013xpa},
and it has been applied to one-dimensional quiver models via a dimensional reduction
\cite{Hwang:2014uwa,Cordova:2014oxa,Hori:2014tda}.
Here, we derive the index of the supersymmetric quiver quantum mechanics
from the beginning in the unpolished way.
We reproduce the Poincar\'{e} polynomials\footnote{
Precisely speaking, the partition function of the supersymmetric quantum mechanics generally gives the $\chi_y$-genus of the moduli space
 \cite{Cordova:2014oxa,Hori:2014tda}.
 However, in examples treating in this paper, the $\chi_y$-genus and the Poincar\'{e} polynomial coincide
 with each other under an identification of parameters. So we do not distinguish the terminology between them
 if there is no misunderstanding.}
and wall crossing formal{\ae} for the quiver moduli spaces even when the corresponding quiver diagrams only include non-Abelian nodes.
Our concrete constructions of the quiver quantum mechanics 
might also be useful to understand more general matrix quantum mechanics embodying $M$-theory
\cite{Banks:1996vh,Berenstein:2002jq}.


The localization, which we utilize throughout the paper, reduces the infinitely many dimensional path integral of the partition function to 
a finite dimensional contour integral with respect to the eigenvalues of an adjoint scalar field.
The localization is powerful tool to understand exactly the non-perturbative effects like dualities
 in the supersymmetric gauge theory.
It can be used in order to derive the non-perturbative (instanton) corrections in $\Ncal=2$ supersymmetric 
Yang-Mills theory \cite{Nekrasov:2002qd,Nekrasov:2003rj}.
In this calculation, the localization fixed points are classified by using the Young diagrams.
The fixed point sets in our results are also interpreted in terms of the attractive combinatorial objects. 

The organization of this paper is as follows:
In section 2, we construct the supersymmetric quantum mechanics with four supercharges by the dimensional reduction from four dimensional $\Ncal=1$ supersymmetric gauge theory. We construct the BRST charge from a linear combination of the four supercharges and redefine the fields of the theory. For the localization to work, we introduce    background gauge fields of R-symmetries, which are also necessary to obtain the partition function in the form of the refined index. We also introduce the  physical observables which can be exactly evaluated by the localization. 
In section 3, we generalize the formulation to the general quiver gauge theory.
In section 4, we derive the exact formul{\ae} for the partition function and observable of the quiver quantum mechanics by using the localization. We also introduce the equivalent $T$-character, which is a convenient tool to calculate the residue integrals.
In section 5, we give some examples of the quiver quantum mechanics by restricting the gauge groups to the Abelian ones and compute the partition functions and the expectation values of the Wilson loop operators. We find that the dependences of the partition functions on the background 
gauge fields of the R-symmetries for the chiral multiplets surprisingly disappear, and our results  perfectly agree with the Poincar\'{e} polynomials
 of the Higgs branch moduli spaces and the wall crossing formal{\ae}.
 On the other hand, the expectation values of the Wilson loop operators depend on those background gauge fields and become rather complicated polynomials.
In section 6, we treat more general cases, which include the non-Abelian gauge groups. By choosing  appropriate  fixed points carefully,  we find that our partition functions again agree with the Poincar\'{e} polynomials
($\chi_y$-genus) and the wall crossing formal{\ae}. 
Especially, we find that when the ranks of the gauge groups are non-coprime, the  fixed points of the vector multiplets do contribute.
The last section is devoted to a summary and future problems.
In the appendix \ref{sec:app}, we give a few examples of the wall crossing formal{\ae} in a special case. In the appendix \ref{mod}, we give an explanation for deformation of Fayet-Iliopoulos parameters to obtain the correct partition function when the ranks of the gauge groups are non-coprime.

\section{$\Ncal=4$ $U(N)$ supersymmetric quantum mechanics}\label{sec:sqm}

We first describe an $\Ncal=4$ $U(N)$ supersymmetric quantum mechanics, which is obtained from the dimensional reduction of  four dimensional $\Ncal=1$ $U(N)$ supersymmetric gauge theory to one dimension.
Four dimensional supersymmetric theory originally has $U(1)$ R-symmetry. After the dimensional reduction, a part of the Lorentz symmetry becomes the R-symmetry of the
reduced theory.
Our supersymmetric matrix quantum mechanics possesses the $SU(2)_J\times U(1)_R$ global R-symmetries
\cite{Cordova:2014oxa}.
We derive the supersymmetric matrix quantum mechanics explicitly for a vector multiplet part and a chiral multiplet part separately by the dimensional reduction. 
We follow the conventions used in \cite{Wess:1992cp} in what follows.

\subsection{Vector multiplet}

In one dimensional $\Ncal = 4$ supersymmetric theory, a vector multiplet is composed of a gauge field $A_0$, three real scalars $X_i$ $(i=1,2,3)$, two complex fermions $\lambda_{\alpha}$ $(\alpha=1,2)$, and an  auxiliary real scalar $D$.  All fields are in the adjoint representation of $U(N)$.
The representations of $SU(2)_J$ and   $U(1)_R$ charges of the vector multiplet are summarized in Table \ref{tab:rv}.

\begin{table}[h]
\centering
\begin{tabular}{c||c|c|c|c} 
~& $A_0$ & $X_i$ & $\lambda_{\alpha}$ & $D$ \\ \hline
$SU(2)_J$ & $\mathbf{1}$ & $\mathbf{3}$ & $\mathbf{2}$ & $\mathbf{1}$ \\ \hline
$U(1)_R$ & 0 & 0 & $\frac{1}{2}$ & 0 \\
\end{tabular} 
\caption{The R-symmetries of the vector multiplet.} 
\label{tab:rv}
\end{table}

Using these fields, the action is given by
\begin{align}
 S_V&=\frac{1}{g^2}\int dt \, \Tr\bigg[\frac{1}{2}(\Dcal_0X_i)^2+\frac{1}{4}[X_i,X_j]^2
 -i\lambdabar \sigmabar^0 \Dcal_0 \lambda +\lambdabar \sigmabar^i [X_i,\lambda]+\frac{1}{2}D^2 - g^2\zeta D\bigg],
 \label{vector action}
\end{align}
where $g$ is the gauge coupling, $\zeta$ is the Fayet-Iliopoulos (FI) parameter, and
\begin{align}
\Dcal_0=\del_0+i[A_0,\cdot],
\end{align}
is the covariant derivative.
The action (\ref{vector action}) is invariant under the following 
supersymmetric transformations:
\be
\begin{array}{lcl}
\delta A_{0} &=& -i\xi \sigma^{0} \lambdabar +i\lambda \sigma^0 \xibar, \\
\delta X_{i} &=& i\xi \sigma^{i} \lambdabar -i\lambda \sigma^i \xibar, \\
\delta \lambda &=& i\xi D +2 \sigma^{0i}\xi \Dcal_0 X_i+i\sigma^{ij}\xi [X_i,X_j] , \\
\delta D &=& -\xi \sigma^0 \Dcal_0 \lambdabar -i\xi \sigma^i [X_i, \lambdabar]
-\Dcal_0\lambda \sigma^0 \xibar -i[X_i,\lambda]\sigma^i \xibar,
\end{array}
\ee
where  $\xi_\alpha$ represents the supersymmetric parameters,
and in terms of the supercharges $Q_\alpha$, the supersymmetric variation is given by
\begin{align}
\delta=\xi Q+\xibar \Qbar.
\end{align}

To apply the localization method to the theory, we now introduce a linear combination of the supercharges by
\begin{align}
 Q&=\frac{i}{\sqrt{2}}(Q^1-\Qbar^1).
\end{align}
We call this the BRST charge conventionally in the following.
We also should consider the Euclidean theory by the Wick rotation $t\to -i\tau$
and define linear combinations of the bosonic fields by
\be
\begin{array}{llll}
Z=X_1-iX_2, & \Zbar=X_1+iX_2,&
\sigma=X_3,& A= A_\tau,\\
Y_{\R}=D-\frac{1}{2}[Z,\Zbar],
\end{array}
\ee
and
\be
\begin{array}{lll}
\lambda_z=\sqrt{2}i\lambdabar_2, & \lambda_{\zbar}=-\sqrt{2}i\lambda_2, & \eta =-\frac{1}{\sqrt{2}}(\lambda_1+\lambdabar_1), \\
\chi_\R =\frac{i}{\sqrt{2}}(\lambda_1-\lambdabar_1),
\end{array}
\ee
for the corresponding fermionic fields, in order to make the localization work well.
This operation is usually called  ``topological twist'', but this is nothing but the redefinitions of the supercharges and fields and
does not change the original theory in one dimension.

Under the introduced BRST symmetry, the fields are transformed by
\be
\begin{array}{lcl}
QZ=i\lambda_{z}, &&Q\lambda_{z}=i(\Dcal_{\tau}Z+[\sigma ,Z]), \\
Q\Zbar=-i\lambda_{\zbar}, && Q\lambda_{\zbar}=-i(\Dcal_{\tau}\Zbar +[\sigma,\Zbar]),\\
QA=i\eta, &&\\
Q\sigma=\eta, && Q\eta=-\Dcal_{\tau}\sigma,\\
QY_{\R}=i(\Dcal_{\tau} \chi_{\R} +[\sigma, \chi_{\R}]), && Q\chi_{\R}=iY_{\R}.
\end{array}
\ee
The BRST transformations are nilpotent up to the time translation and (complexified) gauge transformation.

The Euclidean action of the theory (\ref{vector action}) is written as a $Q$-exact form:
\be
S_V=\frac{1}{2g^2}Q\int d\tau \Tr \bigg[\frac{1}{2}\lambda_{z} \overline{Q\lambda_{z}}+\frac{1}{2}\lambda_{\zbar}\overline{Q\lambda_{\zbar}} +\eta \overline{Q\eta} -\chi_{\R} \overline{Q\chi_{\R}} -2i\chi_{\R}\mu_{\R}\bigg],
\label{Q-exact action}
\ee
where $\mu_{\R}=\frac{1}{2}[Z,\Zbar]-g^2\zeta$ is a (real) moment map constraint which contains the original D-term constraint and describes the moduli space of the vacua.
After integrating out the auxiliary field $Y_\R$, we obtain the Euclidean action of the original matrix quantum mechanics.

The field redefinitions (topological twist) spoil the original R-symmetries,
but the theory is still invariant under the following twisted ``R-transformation" $U(1)_J'$, which acts on the fields by
\be
\begin{array}{lcl}
Z\to e^{i\theta_J}Z, && \lambda_z\to  e^{i\theta_J}\lambda_z,
\end{array}
\ee
with an R-transformation parameter $\theta_J$.
To obtain the refined index, we need a ``gauging'' of this global R-symmetry, which modifies the moduli space of the theory by induced mass terms. 
Under the gauged $U(1)_J'$ symmetry with a constant background $A_J=\e$, the $\tau$-derivatives of $Z$ and $\lambda_z$ are modified into
\begin{align}
\del_{\tau}Z&\to (\del_{\tau}+i\e)Z, ~~~~~~~~~~\del_{\tau}\lambda_z\to (\del_{\tau}+i\e)\lambda_z.
\end{align}
This is known to the $\Omega$-background \cite{Nekrasov:2002qd,Nekrasov:2003rj}.
Thus, we obtain the modified BRST transformations:
\be
\begin{array}{lcl}
Q_\e Z=i\lambda_{z}, &&Q_\e \lambda_{z}=i(\Dcal_{\tau}Z+[\sigma ,Z]+i\e Z), \\
Q_\e\Zbar=-i\lambda_{\zbar}, && Q_\e \lambda_{\zbar}=-i(\Dcal_{\tau}\Zbar +[\sigma,\Zbar]-i\e \Zbar),\\
Q_\e A=i\eta, &&\\
Q_\e \sigma=\eta, && Q_\e \eta=-\Dcal_{\tau}\sigma,\\
Q_\e Y_{\R}=i(\Dcal_{\tau} \chi_{\R} +[\sigma, \chi_{\R}]), && Q_\e \chi_{\R}=iY_{\R}.
\end{array}
\label{BRST transformations}
\ee
The BRST transformations are nilpotent up to the time translation, gauge transformation including the gauged $U(1)_J'$ transformation.

The action of the modified theory is obtained  
by replacing simply $Q$ with $Q_\e$ in (\ref{Q-exact action}).

\subsection{Chiral multiplet}

Let us now construct the theory which includes chiral multiplets.
The chiral multiplet is composed of a complex scalar $q$, two complex fermions $\psi_\alpha$, and an auxiliary complex scalar $F$. The chiral multiplet can be taken in an arbitrary representation of $U(N)$. Here, we take it in the fundamental representation for simplicity.  (A generalization is straightforward.) 
The representations under $SU(2)_J$ and  charges under $U(1)_R$ of the chiral multiplet are summarized in Table \ref{tab:rc}.

\begin{table}[h]
\centering
\begin{tabular}{c||c|c|c} 
~& $q $ & $\psi_{\alpha} $ & $F$ \\ \hline
$SU(2)_J$ & $\mathbf{1}$ & $\mathbf{2}$ &  $\mathbf{1}$ \\ \hline
$U(1)_R$ & $r$ & $r-\frac{1}{2}$ & $r-1$  \\ 
\end{tabular} 
\caption{The R-symmetries of the chiral multiplet with a  $U(1)_R$ charge $r$.}   
\label{tab:rc}
\end{table}　

The action  is given by
\begin{align}
S_C&=\int dt \,  \Tr \bigg[|\Dcal_0 q|^2 -|X_i q|^2  -i\psibar \sigmabar^0 \Dcal_0 \psi +\psibar \sigmabar^i X_i \psi 
 +|F|^2 +i\sqrt{2}(\qbar \lambda\psi  - \psibar \lambdabar q ) +\qbar D q\bigg].
\end{align}
This action is invariant under the following supersymmetric transformations:
\be
\begin{array}{lcl}
\delta q&=&\sqrt{2}\xi \psi, \\
\delta \psi&=&i\sqrt{2} (\sigma^0 \xibar \Dcal_0 q+i\sigma^i \xibar X_iq)+\sqrt{2}\xi F, \\
\delta F&=& i\sqrt{2}  (\xibar\sigmabar^0\Dcal_0\psi +i \xibar\sigmabar^i X_i \psi)+2i\xibar \lambdabar q.
\end{array}
\ee

After the Wick rotation,  we define the bosonic fields:
\be
\begin{array}{lcl}
Y_\C =F+Zq, && \Ybar_\C=\Fbar+\qbar \Zbar, 
\end{array}
\ee
and the fermionic fields:
\be
\begin{array}{lcl}
\psi =\psi_2, && \psibar=\psibar_2,\\
\chi_\C =-\psi_1, && \chibar_\C=-\psibar_1.
\end{array}
\ee
These fields transform under the BRST symmetry by
\be
\begin{array}{lcl}
Qq =i\psi,&& Q\psi=i(\Dcal_{\tau} q+\sigma q),\\
Q\qbar =-i\psibar, && Q\psibar=-i(\Dcal_{\tau}\qbar-\qbar \sigma),\\
QY_{\C} =i(\Dcal_{\tau}\chi_{\C}+\sigma \chi_{\C}), && Q\chi_{\C,}=iY_{\C},\\
Q\Ybar_{\C} =i(\Dcal_{\tau}\chibar_{\C}-\chibar_{\C}\sigma), && Q\chibar_{\C}=i\Ybar_{\C}.
\end{array}
\ee
Using the BRST charge, the Euclidean action can be written as the $Q$-exact form:
\be
S_C=\frac{1}{2}Q \int d\tau \, \Tr \bigg[\psi\overline{Q\psi}+\psibar \overline{Q\psibar}
-\chi_{\C}\overline{Q\chi_{\C}}-\chibar_{\C} \overline{Q\chibar_{\C}} -2i\chibar_{\C} \mu_{\C}-2i\chi_{\C}\mubar_{\C}\bigg],
\ee
where
\be
\begin{array}{lcl}
\mu_{\C} &=& Zq-\frac{\del \Wbar(\qbar)}{\del \qbar}, \\
\mubar_{\C} &=& \qbar \Zbar -\frac{\del W(q)}{\del q},
\end{array}
\label{complex moment maps}
\ee
are (complex) moment map constraints associated with the F-term constraints and $W(q)$ is the superpotential.
By including the chiral multiplet, the real moment map (D-term constraints) also are modified to
\be
\mu_{\R} = \frac{1}{2}[Z,\Zbar] + g^2 (q \qbar -\zeta).
\label{real moment maps}
\ee

After these redefinitions of the fields in the chiral multiplet, the theory possesses the following twisted R-transformations $U(1)_J'\times U(1)_R'$:
\be
\begin{array}{lcl}
q \to e^{ir\theta_R}q, && \psi\to e^{ir\theta_R}\psi, \\
Y_\C \to e^{i(\theta_J+r\theta_R)}Y_\C, && \chi_\C\to e^{i(\theta_J+r\theta_R)}\chi_\C,
\end{array}
\ee
where $\theta_{J,R}$ are the R-transformation parameters.
As similar as the previous section, we gauge these R-symmetries in the constant backgrounds $A_J=\e$ and $A_R=\et$.
The BRST transformations are deformed by
\be
\begin{array}{lcl}
Q_\e q =i\psi,&& Q_\e \psi=i(\Dcal_{\tau} q+\sigma q +ir\et q),\\
Q_\e \qbar =-i\psibar, && Q_\e \psibar=-i(\Dcal_{\tau}\qbar-\qbar \sigma -ir\et \qbar),\\
Q_\e Y_{\C} =i(\Dcal_{\tau}\chi_{\C}+\sigma \chi_{\C}+i(\e+r\et)\chi_{\C}), && Q_\e \chi_{\C,}=iY_{\C},\\
Q_\e \Ybar_{\C} =i(\Dcal_{\tau}\chibar_{\C}-\chibar_{\C}\sigma-i(\e+r\et)\chibar_{\C}), && Q_\e \chibar_{\C}=i\Ybar_{\C}.
\end{array}
\ee
Here, we have used the same symbol $Q_\e$ as that in (\ref{BRST transformations}), but $Q_\e$ is regarded as including the whole gauged  $U(1)_J'\times U(1)_R'$ symmetries in the following.
Thus, the BRST transformations  are now nilpotent up to the time translation, gauge transformation including gauged $U(1)_J'\times U(1)'_R$ transformations.

To summarize, the total action including the chiral multiples is written by a sum of $S_V$ and $S_C$, which are $Q_\e$-exact.
So we can apply the localization arguments (coupling independence) to our models with respect to the BRST charge $Q_\e$.


\subsection{Physical observables}

If an operator ${\cal O}$ is $Q_\e$-closed such that $Q_\e{\cal O}=0$, but not $Q_\e$-exact such that ${\cal O}\neq Q_\e {\cal O}'$,
the vacuum expectation value (vev) of the operator ${\cal O}$ can be evaluated exactly by the localization without changing the coupling
independent property. So the operator ${\cal O}$ belonging to
the $Q_\e$-cohomology (equivariant cohomology) becomes a non-trivial physical observable.
We here discuss the possible observables in the matrix quantum mechanics.

First of all, noting that a combination of the fields
\be
\Phi \equiv \sigma+iA,
\ee
is $Q_\e$-closed itself because of the BRST transformations (\ref{BRST transformations}),
we find that any gauge invariant function of $\Phi$ becomes the physical observable.
A possible gauge invariant function made of $\Phi$ is a supersymmetric
Wilson (Polyakov) loop operator along the Euclidean ``time'' direction:
\be
W_R(\Phi) \equiv \Tr_R {\cal P} \exp{\left\{ i \int d\tau \, (A - i\sigma)\right\}}
=\Tr_R {\cal P} \exp{\left\{  \int d\tau \, \Phi\right\}},
\ee
where ${\cal P}$ stands for the path ordered product and the trace is taken over the representation $R$.
We can evaluate the vev of $W_R(\Phi)$ exactly by using the localization in principle.

Secondly, another interesting physical observable is obtained from the dimensional reduction of the
supersymmetric Chern-Simons (CS) action in three dimensions or BF action in two dimensions,
which explains why lower dimensional gauge theories are exactly solvable \cite{Witten:1992xu,Blau:1993hj,Kapustin:2009kz,Ohta:2012ev}.
In one dimensional model, dimensionally reduced CS type operator becomes
\be
{\cal O}_V = \int d\tau\,
\Tr \left\{
\Zbar \Dcal_\tau Z+\Zbar [\sigma,Z] + i\e \Zbar Z
+\lambda_z \lambda_{\zbar}
\right\}.
\ee
We can check this operator is $Q_\e$-closed from the BRST transformations (\ref{BRST transformations}).
We can also define a similar $Q_\e$-closed operator from the chiral multiplet as follows:
\be
{\cal O}_C = \int d\tau\,
\Tr \left\{
\qbar \Dcal_\tau q+\qbar \sigma q + ir\et \qbar q
+\psi \psibar
\right\}.
\ee 
Using the similar arguments in \cite{Witten:1992xu}, we expect that the vev of
the $Q_\e$-closed operator $\exp\left\{-t_V {\cal O}_V -t_C {\cal O}_C \right\}$
gives a partition function of a bosonic (non-supersymmetric) matrix quantum mechanics,
since
\be
\begin{split}
\left\langle e^{-{\cal O}_V -{\cal O}_C} \right\rangle
&=\int \Dcal (\text{bosons}) \Dcal (\text{fermions}) \, e^{- {\cal O}_V - {\cal O}_C} e^{-S_V-S_C}\\
&=\int \Dcal \sigma \Dcal^2 Z \Dcal^2 q \Dcal^2 \lambda_z \Dcal^2 \psi \, e^{-{\cal O}_V -{\cal O}_C}\\
&=\int \Dcal \sigma \Dcal^2 Z \Dcal^2 q \,
e^{- S_\text{MQM}},
\end{split}
\ee
where
\be
S_\text{MQM} = \int d\tau\,
\Tr \left\{
\Zbar \Dcal_\tau Z+\Zbar [\sigma,Z] + i\e \Zbar Z
+\qbar \Dcal_\tau q+\qbar \sigma q + ir\et \qbar q
\right\}.
\ee
Here we have used the coupling independence of the $Q_\e$-exact action and performed the Gaussian
integrals of the fermions.

We will discuss the exact vevs of the above physical observables later.

\section{Quiver quantum mechanics}

\begin{figure}[htbp]
 \begin{center}
\begin{tikzpicture}
  \node (N1) {\circled{$N_1$}};
  \node (N2) [right of=N1] {\circled{$N_2$}};
  \node (N3) [below of=N2] {\circled{$N_3$}};
  \node (N4) [right of=N3] {\circled{$N_4$}};
  \node (N5) [node distance=2.0cm, right of=N4, above of=N4] {\circled{$N_5$}};
  \node (N6) [right of=N5] {\circled{$N_6$}};
  \draw[->] (N1) to node {$k_1$} (N2);
  \draw[->] (N2) to node [swap] {$k_2$} (N3);
  \draw[->] (N3) to node [swap] {$k_3$} (N4);
  \draw[->] (N4) to node {$k_4$} (N2);
  \draw[->] (N4) to node {$k_5$} (N5);
  \draw[->] (N6) to node {$k_6$} (N5);
\end{tikzpicture}
 \end{center}
 \caption{A generic quiver diagram with 6 nodes.}
 \label{quiver diagram}
\end{figure}
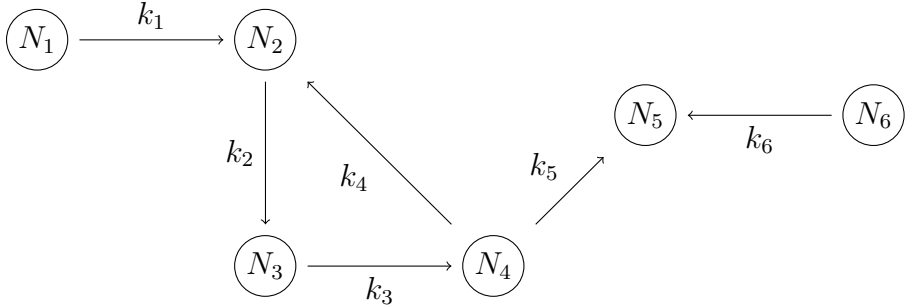

Let us now apply the general formulation of the matrix quantum mechanics in the previous section
to the quiver quantum mechanics.
The quiver gauge theory has a gauge symmetry of a product of gauge groups $G=\prod_v G_v$
and contains chiral matter multiplets represented between two gauge groups.
We assume all gauge groups are unitary groups of rank $N_v$, that is, $G_v=U(N_v)$.
The structure of the quiver gauge theory 
is represented by the so-called quiver diagram depicted in Figure \ref{quiver diagram}. A quiver diagram is composed of nodes $v$ and arrows $a$ whose ends are attached on the nodes. A node $v$ represents a vector multiplet in the adjoint representation of $U(N_v)$. An arrow, whose direction is $v\to w$, represents a chiral multiplet in the bi-fundamental representation $(\square,\bar{\square})$ of $U(N_v)\times U(N_w)$. If ends of an arrow are attached on the same node, the  chiral multiplet is in the adjoint representation.

Following the representations of the quiver quantum mechanics,
we see the BRST transformations for the vector multiplet at the node $v$ are
\be
\begin{array}{lcl}
Q_\e Z_v = i\lambda_{z,v}, && Q_\e \lambda_{z,v} = i(\del_{\tau}Z_v+[\Phi_v ,Z_v]+i\e Z_v),\\
Q_\e \Zbar_v = -i\lambda_{\zbar,v}, && Q_\e \lambda_{\zbar,v} = -i(\del_{\tau}\Zbar_v +[\Phi_v,\Zbar_v]-i\e \Zbar_v), \\
Q_\e A_{v} = i\eta_v, && \\
Q_\e \sigma_v = \eta_v, && Q_\e \eta_v = -(\del_{\tau}\sigma_v+[\Phi_v,\sigma_v]),\\
Q_\e Y_{\R,v} = i(\del_{\tau} \chi_{\R,v} +[\Phi_v, \chi_{\R,v}]), && Q_\e \chi_{\R,v} = iY_{\R,v},
\label{BRST vector}
\end{array}
\ee
where $\Phi_v \equiv \sigma_v  +i A_{v}$,
and 
the BRST transformations for the chiral multiplet along the arrow $a:v\to w$ are
\be
\begin{array}{lcl}
Q_\e q_a = i\psi_a, && Q_\e \psi_a = i(\del_{\tau} q_a+\Phi_v q_a-q_a\Phi_w +i\e_a q_a),\\
Q_\e \qbar_a = -i\psibar_a, && Q_\e \psibar_a = -i(\del_{\tau}\qbar_a-\qbar_a \Phi_v+\Phi_w \qbar_a-i\e_a \qbar_a),\\
Q_\e Y_{\C,a} = i(\del_{\tau}\chi_{\C,a}+\Phi_v \chi_{\C,a}- \chi_{\C,a}\Phi_w+i(\e+\e_a)\chi_{\C,a}),
&& Q_\e \chi_{\C,a} = iY_{\C,a},\\
Q_\e \Ybar_{\C,a} = i(\del_{\tau}\chibar_{\C,a}-\chibar_{\C,a}\Phi_v+\Phi_w\chibar_{\C,a}-i(\e+\e_a)\chibar_{\C,a}),
&& Q_\e \chibar_{\C,a} = i\Ybar_{\C,a},
\end{array}
\label{BRST chiral}
\ee
where we have defined
\begin{align}
\e_a\equiv r_a \et.
\end{align}
If there is no superpotential, the $U(1)_R'$ symmetry exists independently in each chiral multiplet. Thus, $\e_a$ is interpreted as a constant  $U(1)_R'$ gauge field which is coupled to the chiral multiplet labeled by $a$ with unit charge.

For later convenience, we introduce a vector notation of the fields.
For bosonic and fermionic fields in the vector multiplet at each node $v$, we define
$\vec{\Bcal}_v\equiv (Z_v,\Zbar_v,\Phibar_v,Y_{\R,v})$
and $\vec{\Fcal}_v\equiv (\lambda_{z,v},\lambda_{\zbar,v},\eta_v,\chi_{\R,v})$, respectively, where $\Phibar_v \equiv
\sigma_v-iA_v$.
For the chiral multiplets, we define $\vec{\Bcal}_a\equiv (q_a,\qbar_a,Y_{\C,a},\Ybar_{\C,a})$
and $\vec{\Fcal}_a\equiv (\psi_{a},\psibar_{a},\chi_{\C,a},\chibar_{\C,a})$.
The total action is
\begin{align}
S&=S_V+S_C,
\end{align}
where
\begin{align}
S_V&=\frac{1}{2g^2}Q_\e\int d\tau \, \sum_v\Tr \bigg[\vec{\Fcal}_v\cdot\overline{Q_\e\vec{\Fcal}_v}
 -2i\chi_{\R,v}\mu_{\R,v}\bigg],
\end{align}
and
\begin{align}
S_C&=\frac{1}{2}Q_\e \int d\tau \, \sum_a \Tr \bigg[\vec{\Fcal}_a\cdot\overline{Q_\e\vec{\Fcal}_a}-2i\chibar_{\C,a} \mu_{\C,a}-2i\chi_{\C,a}\mubar_{\C,a}\bigg].
\end{align}
Here ``$\cdot$'' denotes an inner product of the vectors of the fields with a suitable norm. 
The moment map constraints are now given by
\begin{align}
\mu_{\R,v}&=\frac{1}{2}[Z_v,\Zbar_v] + g^2 \bigg(\sum_{a:v \to \bullet}q_a \qbar_a-\sum_{a:\bullet\to v}\qbar_aq_a -\zeta_v\bigg),
\label{D-term equations} \\
\mu_{\C,a}&=Z_vq_a-q_aZ_w-\frac{\del \Wbar(\qbar)}{\del \qbar_a},~~~~~\mubar_{\C,a}=\qbar_a \Zbar_v-\Zbar_w \qbar_a -\frac{\del W(q)}{\del q_a},
\end{align}
where $\bullet$ denotes all nodes which are connected to $v$ by arrows in the quiver diagram,
and the FI parameters $\zeta_v$ should satisfy $\theta({\bf N}) \equiv \sum_v N_v \zeta_v=0$ \cite{Denef:2002ru}.
If there is no closed oriented loop in the quiver diagram, the corresponding quiver gauge theory cannot possess the superpotential by the gauge invariance.

The physical observables discussed in the previous section are generalized to
the quiver theory. We can naively define a  $Q_\e$-closed Wilson loop operator
as a product of Wilson loop operators with respect to each node $v$:
\be
\prod_v W_{R_v}(\Phi_v).
\ee
However, as we will see, the above Wilson loop generally contains the diagonal $U(1)$ part
of the quiver group, which associated with the center of mass integral coordinate.
If the Wilson loop has an overall factor like $e^{i\int d\tau \phi_c}$,
where $\phi_c$ is the center of mass coordinate,
the vev vanishes after integrating over the center of mass coordinate.
So we should use the center of mass free Wilson loop operators, which
are written typically in terms of the ``relative'' coordinates associated with the arrows
$a:v\to w$
\be
W_{R_v}(\Phi_v)W_{R_w}(-\Phi_w)=\Tr_{R_v}{\cal P}e^{\int d\tau\, \Phi_v}\, \Tr_{R_w}{\cal P}e^{-\int d\tau\, \Phi_w},
\ee
where $R_v$ and $R_w$ are suitably chosen to remove the center of mass coordinate.
Then, we can evaluate the non-vanishing vev for the Wilson loop operator.

In addition, the sum of the CS-type action,
\be
\sum_v {\cal O}_{V,v} + \sum_a {\cal O}_{C,a},
\ee
where
\begin{align}
{\cal O}_{V,v} &= \int d\tau\,
\Tr \left\{
\Zbar_v \Dcal_\tau Z_v+\Zbar [\sigma_v,Z_v] + i\e \Zbar_v Z_v
+\lambda_{z,v} \lambda_{\zbar,v}
\right\},\\
{\cal O}_{C,a} &= \int d\tau\,
\Tr \left\{
\qbar_a \Dcal_\tau q_a+\qbar_a \sigma q_a + ir\et \qbar_a q_a
+\psi_a \psibar_a
\right\},
\end{align}
 is $Q_\e$-closed in the quiver quantum mechanics.
We can derive the partition function of a bosonic quiver quantum mechanics
as the vev of the above CS -type $Q_\e$-closed operator.

\section{Exact partition function of quiver quantum mechanics}

Using the formulation and field redefinitions in the previous sections, 
we derive a generic formula for the partition function of the quiver quantum mechanics exactly
by the localization method.
In order to find the relation between the exact partition function and (refined) index of the
BPS states,
we assume that the Euclidean time direction is compact as $\tau \sim \tau +\beta$
in the following calculations.

\subsection{Localization}

As we have seen in section \ref{sec:sqm}, the reduced supersymmetric Yang-Mills action is written as an exact form
of a part of the supercharges.
If we denote by $\Bcal^I$ and $\Fcal^I$ the vectors of the bosonic and fermionic fields, respectively, we find that a $Q$-exact supersymmetric Yang-Mills action takes the following form generically:
\begin{align}
S&=t Q \int d\tau \, \Tr \big[ g_{IJ}\Fcal^I \overline{Q\Fcal^J}\big]-2it' Q \int d\tau \, \Tr \chi_i\mu^i,
\label{generic action}
\end{align}
where we have introduced the different coupling constants $t$ and  $t'$ for the Gaussian and constraint part.
A metric $g_{IJ}$ gives a norm on the field variables and $\chi^i$ are superpartners of the auxiliary fields $Y^i$,
which will  be the moment map constraints $\mu^i=0$. Note that there exists a $Q$-closed combination of the fields
$\Phi$ in addition to $\Bcal^I$ and $\Fcal^I$.

We can show that the partition function for the generic $Q$-exact action (\ref{generic action})
\be
\Zcal = \int \Dcal \Phi \prod \Dcal \Bcal^I \prod \Dcal  \Fcal^I \,
e^{-S},
\ee
or the vev of the $Q$-closed operator
\be
\left\langle {\cal O} \right\rangle
=\frac{1}{Z}\int \Dcal \Phi \prod \Dcal \Bcal^I \prod \Dcal  \Fcal^I \, {\cal O} \,
e^{-S},
\ee
is independent of the couplings $t$ and $t'$.
So we can take the limit of $t,t'\to \infty$ (weak coupling limit) without changing the value of the
partition function or vev.

If we first take the limit of $t\to \infty$, we find that the path integral becomes WKB exact
and localizes at the fixed points  $Q\Fcal^I=Q\Bcal^I=0$,
since the action with the coupling $t$ is essentially Gaussian.
The Gaussian integral also induces Jacobians (1-loop determinants) to the measure, which are given by
super determinants (super Hessian) of the BRST transformations
evaluated at the fixed points \cite{Karki:1993bw,Bruzzo:2002xf,Bruzzo:2003rw,Ohta:2012ev,Miyake:2011yr,Ohta:2013zna}
\begin{align}
\Delta(\Phi)&=
\left.
\sqrt{ \frac{\det \frac{\delta Q\Bcal^I}{\delta \Fcal^J}}{\det \frac{\delta Q\Fcal^I}{\delta \Bcal^J}}}\right|_{Q\Fcal^I=Q\Bcal^I=0}. \label{eq:detformula}
\end{align}
The dependence of the coupling $t$ disappears from the 1-loop determinants as expected,
because of the cancellation between the bosons and fermions.



On the other hand, if we take the limit of $t'\to \infty$, the action with the coupling $t'$ imposes
 delta-functional constrains $\mu^i=0$ in the path integral after integrating out the auxiliary field $Y^i$.
These constraints work in any coupling region because of the coupling independence.
So we always should take the moment map constraints $\mu^i=0$ into account in addition to the fixed points.

To summarize, the partition function or vev of theory with $Q$-exact action is given by a summation over
a finite set of the fixed points on the moment map constraints (vacuum moduli space).
We obtain the Duistmaat-Heckman localization formula for supersymmetric Yang-Mills theory
\begin{align}
Z &= \sum_{\Phi^* \in \text{fixed points}} \Delta_{gh}(\Phi^*)\Delta(\Phi^*),\\
\left\langle {\cal O} \right\rangle
&=\frac{1}{Z}\sum_{\Phi^* \in \text{fixed points}} \Delta_{gh}(\Phi^*)\Delta(\Phi^*) {\cal O},
\end{align}
where $\Delta_{gh}(\Phi^*)$ is the 1-loop determinant from the Faddeev-Popov ghosts, which we will discuss
in the next subsection.
Our residual task is to find the explicit 1-loop determinants and solutions of the fixed points of
the quiver quantum mechanics.

\subsection{Gauge fixing}

Originally, the quiver quantum mechanics has the gauge symmetry of $\prod_v U(N_v)$,
but the existence of the real adjoint scalar fields $\sigma_v$ enhances the gauge symmetry
to the complexified one, namely $\prod_v GL(N_v,\C)$.
We can regard a $Q_\e$-closed combination of the gauge field and adjoint scalar $\Phi_v=\sigma_v + iA_{v}$
as a holomorphic section of a $GL(N_v,\C)$ factor.
Under these complexified gauge symmetries, the D-term constraints ($\mu_{\R,v}=0$) become redundant,
since there is an isomorphism in the vacuum moduli space given by the moment map quotient space
\be
{\cal M} \equiv \frac{\mu_\R^{-1}(0)\cap \mu_\C^{-1}(0) \cap  \bar{\mu}_\C^{-1}(0)}{U(N)}
\simeq
\frac{\mu_\C^{-1}(0) \cap  \bar{\mu}_\C^{-1}(0)}{GL(N,\C)}.
\ee
So we can utilize these complexified gauge symmetries instead of taking D-term constraints into account directly.

Using the complexified gauge symmetries, we can choose the gauge  condition so that the off-diagonal components of $\Phi_v$ to be zero as similar as the matrix models,
\begin{align}
\Phi_v|_{\text{off-diag}}=0.
\end{align}
After imposing the gauge condition, we still have $GL(1,\C)^N$ gauge symmetries. To fix the residual gauge degrees of freedom, we choose
\begin{align}
A^i_{v}=0, \qquad(i=1,\ldots,N_v),
\end{align}
for the Cartan (diagonal) part of the each gauge field.
In other words, the imaginary part of $\Phi_v$ is chosen to be zero.

At this moment, the total number of the gauge conditions is $2(N^2-N)+N=2N^2-N$.
Then, we still have $N$ gauge degrees of freedom, since $GL(N,\C)$ possesses  $2N^2$ degrees of freedom.
However, as we explained above, since this residual gauge symmetry is equivalent to treat the Abelian part of the D-term conditions, we can fix it by imposing the real moment map constraints on the fixed point sets after the
gauge fixing above. We will see that the solutions to the D-term constraints correspond to a choice of integral contour
over $\Phi_v$. Then the residual gauge degrees of freedom are fixed by the choice of the integral contour.
This is the essential reason why the wall crossing phenomena occurs in the quiver matrix model.
Each solution of the D-term condition (stability condition) relates to the choice of the contour
and the value of the partition function (index) or vev in the quiver
quantum mechanics jumps by each choice of the contour.
We will see this relation in more concrete examples later.

At least, in the above gauge choice, we obtain the gauge fixing action with the Faddeev-Popov ghosts $c_v,\cbar_v$
\begin{align}
S_{\text{gh}}&=i \int d\tau \, \Tr [\cbar_v (\del_\tau c_v +[\Phi_v,c_v])],
\end{align}
for each $U(N_v)$ node.
This gives the 1-loop determinant for the ghosts
\be
\begin{split}
\Delta_{\text{gh},v}(\phi)&=\prod_{n=1}^{\infty}\bigg(\frac{2\pi  n}{\beta}\bigg)^{2N_v}\prod_{\substack{i\neq j}}\prod_{n=-\infty }^{\infty}\bigg(\frac{2\pi i n}{\beta}+\phi_{v,i}-\phi_{v,j}\bigg)\\
&=\beta^{N_v}\prod_{i\neq j}2 \sinh\frac{\beta}{2}(\phi_{v,i}-\phi_{v,j}),
\end{split}
\ee
where $\phi_{v,i}$ ($i=1,\ldots,N_v$) are the diagonal components of $\Phi_v$,
and we have used the infinite product representation of the hyperbolic sine function and zeta-function regularization
for the infinite product.

\subsection{Partition function} \label{sec:part}

Now, let us find the exact partition function of the quiver quantum mechanics.
Under the diagonal gauge, the BRST fixed point equation
becomes
\be
Q_\e \eta_v = -\del_\tau \sigma_v = 0,
\ee
that is, only the constant modes of $\sigma_v$ ($\Phi_v$) survive in the path integral.
These constant eigenvalues of $\Phi_v$ are denoted by $\phi_1,\ldots,\phi_{N_v}$ in what follows.

Using the super determinant formula (\ref{eq:detformula}), we find that the 1-loop determinant for a vector multiplet,
including the gauge fixing ghost contribution, with a node $v$ is 
given by
\begin{align}
\Delta_{v}^V(\phi)
&=\Delta_{\text{gh},v}(\phi) \sqrt{\frac{\det \frac{\delta Q_\e \Bcal_v^I}{\delta \Fcal_v^J}}
{\det \frac{\delta Q_\e \Fcal_v^I}{\delta \Bcal_v^J}}}\\
&=\Delta_{\text{gh},v}(\phi)\prod_{i, j=1}^{N_v}\prod_{n=-\infty}^{\infty}\frac{1}{\frac{2\pi i n}{\beta}+\phi^v_i-\phi^v_j+i\e} \notag \\
&=\bigg(\frac{\beta }{2i\sin \frac{\beta \e }{2}}\bigg)^{N_v}\prod_{\substack{i\neq j}}\frac{\sinh\frac{\beta}{2}(\phi^v_i-\phi^v_j)}{\sinh\frac{\beta}{2}(\phi^v_i-\phi^v_j+i\e)},
\end{align}
in terms of the constant eigenvalues of $\Phi_v$ at the fixed points.
Similarly, the 1-loop determinant for a chiral multiplet with an arrow $a:v\to w$ becomes
\begin{align}
\Delta_{a}^C(\phi)
&=\sqrt{\frac{\det \frac{\delta Q_\e \Bcal_a^I}{\delta \Fcal_a^J}}
{\det \frac{\delta Q_\e \Fcal_a^I}{\delta \Bcal_a^J}}}\\
&=\prod_{i=1}^{N_v}\prod_{i'=1}^{N_w}\prod_{n=-\infty}^{\infty}\frac{\frac{2\pi in}{\beta}+\phi^v_i-\phi_{i'}^w+i(\e+\e_a)}{\frac{2\pi in}{\beta}+\phi^v_i-\phi_{i'}^w+i\e_a} \notag \\
&=\prod_{i=1}^{N_v}\prod_{i'=1}^{N_w}\frac{\sinh\frac{\beta}{2}(\phi_i^v-\phi^w_{i'}+i(\e+\e_a))}{\sinh\frac{\beta}{2}(\phi_i^v-\phi^w_{i'}+i\e_a)}.
\end{align}
We finally obtain an integral formula for the partition function of the generic quiver quantum mechanics
\begin{align}
\Zcal&=\int \prod_v\frac{1}{N_v!}\prod_{i=1}^{N_v}\frac{d\phi^v_i}{2\pi i} \Delta_{v}^V(\phi) \prod_a \Delta_{a}^C(\phi),
\label{integral formula}
\end{align}
where the $1/N_v!$ factor comes from the order of the Weyl group.

We give two comments on the formula of the partition function. Firstly, since the integrand of the partition function depends only on the relative variables $\phi^v_i-\phi^v_j$ or $\phi_i^v-\phi^w_{j'}$, one trivial integration is left, which leads to the infrared divergence. This is due to the center of mass motion of the system \cite{Denef:2002ru}. We will ignore this divergence in the subsequent sections.
Secondly, we note on the contour integrals over the constant modes $\phi_i^v$. This means that there still exists the
gauge degrees of freedom as we have  discussed above.
The integral formula contains poles in the denominator of the integrand.
Choosing a suitable contour, some poles in the denominator are picked up
as the residue integral, which correspond to the BRST fixed points on the moment map constraints.

Thus, we find the localization formula for the partition function of the quiver quantum mechanics:
\be
\Zcal=\sum_{\phi^* \in \text{fixed points}}
\prod_v\frac{1}{N_v!}\prod_{i=1}^{N_v}\Delta_{v}^V(\phi^*) \prod_a \Delta_{a}^C(\phi^*),
\label{eq:formulaz}
\ee
where $\phi^*$ stands for the fixed point sets which satisfy the BRST fixed point equation
and the moment map constraints, and the 1-loop determinants are evaluated at the fixed points
in a sense of the residue integral.

Similarly, the vev of the $Q_\e$-closed operator is also evaluated exactly by the localization:
\be
\left\langle {\cal O}(\Phi) \right\rangle
=\frac{1}{\Zcal}\sum_{\phi^* \in \text{fixed points}}
\prod_v\frac{1}{N_v!}\prod_{i=1}^{N_v}\Delta_{v}^V(\phi^*) \prod_a \Delta_{a}^C(\phi^*)
\, {\cal O}(\phi^*).
\ee

We apply the localization formula to some concrete examples in the subsequent sections.

Before closing this section, we make a comment.
The integrand of the formula (\ref{integral formula}) has infinitely many poles due to the trigonometric property
of the determinant. These infinite poles are related to the large gauge transformations
\begin{align}
\phi_i^v\to \phi^v_i+\frac{2\pi i w^v_i}{\beta}, \qquad (w^v_i\in \mathbb{Z}),
\end{align}
since the imaginary part of $\Phi_v$ corresponds to the gauge field $A_v$ \cite{Cordova:2014oxa}.
Then, a position of a pole moves to the other ones by the large gauge transformations. Thus, the infinitely many poles represent the non-perturbative effects of the theory. However, since the 1-loop determinants and the operators are invariant under the large gauge transformations, the partition function and vevs of the operators trivially diverge if we take into account all of the  poles. To avoid this, we only pick up one of the poles in the hyperbolic sine function in the following sections.

\subsection{Fixed points and character}

After taking the diagonal gauge, the gauge symmetry of the quiver theory breaks to the product of the Abelian gauge groups,
which is a maximal torus group $T$. So we have to find the fixed point set of the BRST equations $Q_\e \vec{\Fcal} = Q_\e \vec{\Bcal}=0$ and
the moment map constraints $\mu_\R=\mu_\C=\bar{\mu}_\C=0$ with respect to the maximal torus. This situation is the same as in the Atiyah-Bott-Berline-Vergne
localization formula. The 1-loop determinants correspond to the equivariant Euler class of the localization formula.

The fixed point set of the torus $T$ is classified by combinatorial objects such as
partitions or Young diagrams, similar to the Nekrasov formula 
of the instanton counting \cite{Nekrasov:2002qd,Nekrasov:2003rj}.
Once we find the complete set of the fixed points, we can evaluate the 1-loop determinants at each fixed point in the sense of the residue integral (\ref{integral formula}), which should be equivalent to the JK residue formula in
\cite{Hwang:2014uwa,Cordova:2014oxa,Hori:2014tda}.
However, in the instanton calculus, there exists more convenient way to evaluate the 1-loop determinant at the fixed points,
that is, the evaluation of the equivariant $T$-character \cite{Nakajima,2003math.....12059M}.
We would like to claim that the equivariant $T$-character is also useful
to evaluate the more complicate residue integral
rather than four-dimensional instanton calculus, in the sense of reduction of the calculation.

The character maps the products in the 1-loop determinant of the partition function to summations of the polynomials.
We schematically give a correspondence between the 1-loop determinants of the partition function and $T$-character as follows:
\begin{center}
\begin{tabular}{c|c}
1-loop determinants &  $T$-character \\
\hline
product &   sum \\
$(1-t)^{\pm 1}$  &   $\pm t$
\end{tabular}
\end{center}
As a result, cancellation of the poles and zeros in the residue integral is suitably managed.
In particular, the sum of the polynomials in the $T$-character is easier to handle in
the computer algebra system like {\it Mathematica}.
This is a technical reason why we have introduced the $T$-character in the localization of
quiver quantum mechanics.\footnote{
As we will see in some examples later, the residue integral formula is still important to find
correct fixed points (pole structure).
}

Using the mapping rules, we find that the $T$-character for the quiver quantum mechanics is given by
\be
{\Tcal}(\Phi) = (1-t) \times \left[
\sum_v V_v\times V_v^* - \sum_{a:v\to w} V_v\times V_w^* \times x_a -1
\right],
\label{eq:tcharacter}
\ee
where we have defined
\be
V_v \equiv \Tr \, e^{\beta \Phi_v}  =\sum_{i=1}^{N_v} u_{v,i}, \quad
V_v^* \equiv \Tr \, e^{-\beta \Phi_v} = \sum_{i=1}^{N_v} u_{v,i}^{-1},
\ee
and
\be
t\equiv e^{i\beta \e}, \quad x_a \equiv e^{i \beta \e_a}.
\ee
We here have inserted ``$-1$'' in the square bracket of (\ref{eq:tcharacter}) in order to remove a singularity which corresponds to the infrared divergence due to the center of mass motion.
Strictly speaking, the value of the partition function obtained from the above $T$-character is
 slightly different from the original residue integral (\ref{eq:formulaz}), but
the essential part of the Higgs branch index still holds as we will see.
We denote the partition function evaluated from the $T$-character by $\hat{\Zcal}$,
which represents a contribution from the relative coordinates,
to be distinguished. Then,
the original partition function $\Zcal$ is related to $\hat{\Zcal}$ by
\be
\Zcal = Ct^{-\frac{1}{2}\left(\sum_{a:v\to w}N_vN_w-\sum_v N_v^2 +1\right)}
\hat{\Zcal}, \label{eq:originalp}
\ee
where $C$
is a renormalized constant of the divergence coming from the integral over the center of mass motion.
We set $C=1$ in the following.
Note here that the exponent $\sum_{a:v\to w}N_vN_w-\sum_v N_v^2 +1$ is the dimension of the Higgs branch 
in the relative coordinates. 

We need to classify the whole set of the fixed points and evaluate the $T$-character at the fixed points,
but it is difficult to explain for the general quiver model. In the following section, we give some concrete examples.
We will explain how to classify the fixed points and evaluate the $T$-character and partition function
for more explicit models.

\section{Abelian nodes} \label{sec:abelian}

In the following two sections, we give several examples of the quiver quantum mechanics and compute the partition functions and the vevs of the physical observables exactly using the localization formula. Here, we treat the theories with no closed oriented loop  in the quiver diagrams for simplicity, but the generalization to the case with closed oriented loops will be straightforward. In this section, we consider the Abelian quiver  quantum mechanics.

\subsection{Two nodes}

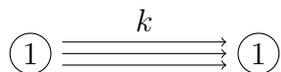
\begin{figure}[htbp]
 \begin{center}
\begin{tikzpicture}
  \node (1) {$\circled{1}$};
  \node (2) [right of=1] {$\circled{1}$};
  \draw[->] (1.20) to node {$k$} (2.160);
  \draw[->] (1) to node {} (2);
  \draw[->] (1.340) to node {} (2.200);
\end{tikzpicture}
 \end{center}
 \caption{$U(1)_1\times U(1)_2$ quiver with $k$ arrows.}
 \label{u1-u1}
\end{figure}

We first start with a simplest example which consists of  two Abelian nodes and $k$ arrows between them,
which is depicted in Fig.~\ref{u1-u1}. In the Higgs picture, this theory describes the BPS bound states of two D-branes  which are wrapped around two distinct cycles in a Calabi-Yau three manifold ($\text{CY}_3$) and intersect transversely in $k$ points \cite{Denef:2002ru}. The adjoint scalars in the  two vector multiplets $X_v^i~(v=1,2)$  represent the positions of the two D-branes in the non-compact space and the  chiral multiplet represents the open string localized at the  intersection point, whose ends are attached on the different D-branes. In the Coulomb picture, this theory describes the BPS bound states of two  particles with mutually nonlocal charges
$\gamma_1$ and $\gamma_2$ in a four dimensional $\Ncal=2$ supergravity theory which arises in the low energy limit of type II string theory compactified on $\text{CY}_3$ \cite{Denef:2002ru}. The number of the chiral multiplets $k$ corresponds to the Dirac-Schwinger-Zwanziger (DSZ) product of the two charges $\langle \gamma_1,\gamma_2\rangle$.

The BRST fixed point equation $Q_\e \psi_a = 0$ says that
\be
q_a(\tau) = h_a e^{-(\phi_1 - \phi_2+i\e_a)\tau},
\ee
where $h_a$ is an integral constant. Forcing the periodic boundary condition $q_a(\tau+\beta)=q_a(\tau)$,
we see that $h_l=0$ except for only one label $a$ ($l\neq a$) by setting
\be
\phi_1 - \phi_2+i\e_a = \frac{2\pi i}{\beta} n, \quad n\in \Z.
\label{u1-u1 solution}
\ee
This  represents the infinitely many poles of the hyperbolic sine function.
But, as we have explained in section \ref{sec:part}, we only pick up a pole with $n=0$.
This fixed point is denoted by $\Phi^*_a$.

The non-vanishing complex scalar $q_a$ at the fixed point is also the solution to the D-term conditions,
\begin{align}
|q_a|^2 &= \zeta_1,\\
-|q_a|^2 &= \zeta_2,
\end{align}
where the FI parameters satisfy the constraint $\theta({\bf N})=\zeta_1+\zeta_2=0$.
The D-term equations mean that there exist the fixed points ($|q_a|\neq 0$)
if $\zeta\equiv\zeta_1=-\zeta_2>0$, but not if $\zeta<0$.
The partition function vanishes for the later case.
This is nothing but the wall crossing formula for the quiver quantum mechanics \cite{Denef:2002ru, Denef:2007vg}.

Let us evaluate the partition function for the case of $\zeta>0$.
The solution (\ref{u1-u1 solution}) means that
\be
u_1 u_2^{-1} x_a = 1,
\label{u1-u1 equation}
\ee
where $u_1=e^{\beta\phi_1}=V_1$, $u_2=e^{\beta\phi_2}=V_2$, and $x_a=e^{i\beta \e_a}$.

The $T$-character of ${\bf N}=(1,1)$ with $k$ arrows becomes\footnote{
We put indices to the $T$-character and partition functions to indicate the dimension vector and the number of arrows.}
\begin{align}
\Tcal^k_{1,1}(\Phi^*_a)&=(1-t)\left(2-\sum_{l=1}^k x_l x_a^{-1}-1\right)\\
&=(t-1)\sum_{l \neq a}  x_l x_a^{-1},
\end{align}
at each fixed point labelled by $a=1,\ldots,k$.
Using the mapping rules between the character and 1-loop determinants
in the partition function, we obtain a contribution from each fixed point to
the partition function
\begin{align}
\Pcal^k_{1,1}(\Phi^*_a) 
&=\prod_{l\neq a}\frac{x_a-tx_l}{x_a-x_l}.
\end{align}
The total partition function is given by a summation over the fixed point set
\be
\begin{split}
\hat{\Zcal}^k_{1,1} &= \sum_{a=1}^k \Pcal^k_{1,1}(\Phi^*_a) \\
&= \frac{1-t^k}{1-t},
\end{split}
\ee
where all $x_a$($\e_a$) dependences surprisingly disappear.

The polynomial of $\hat{\Zcal}_{1,1}^k$ in $t$ is the Poincar\'e polynomial of $\C P^{k-1}$, which is the Higgs branch moduli.
To compare it with the known results in \cite{Denef:2002ru,Manschot:2010qz},
setting $t^{1/2}=-y$, the original partition function becomes
\be
\Zcal^k_{1,1} = (-1)^{k+1}\frac{y^k-y^{-k}}{y-y^{-1}}.
\ee
This result agrees with them.

We can also evaluate the vev of the supersymmetric Wilson loop.
The insertion of the supersymmetric Wilson loop in the path integral does not violate the
above localization arguments. The Wilson loop operator takes a value at the localization fixed point.
At the fixed point, we find a center of mass independent Wilson loop
\be
W(\Phi_a^*) = e^{i\beta (\Phi_1^*-\Phi^*_2)} =x_a^{-1},
\ee
by using the solution to (\ref{u1-u1 equation}).
Thus, we obtain the vev of the Wilson loop by
\be
\begin{split}
\left\langle 
W(\Phi)
\right\rangle
&=\frac{1}{\Zcal^k_{1,1}}
\sum_{a=1}^k
\Pcal^k_{1,1}(\Phi^*_a) x_a^{-1}\\
&=\frac{(1-t)t^{k-1}}{1-t^k}\sum_{a=1}^k x_a^{-1}.
\end{split}
\ee
The dependence of $x_a$ ($\e_a$) does not disappear in contrast to the partition function.

\subsection{$n$ nodes}

Let us now generalize the above result to the Abelian quiver with $n$-nodes.
We assume that  the direction of the arrows between two nodes is identical. Then
the quiver structure is specified by an antisymmetric matrix $K_{ij}=-K_{ji}$ ($i,j=1,\ldots,n$),
where $|K_{ij}|$ represents the number of the arrows from node $i$ to $j$ if $K_{ij}>0$,
or the arrows from node $j$ to $i$ if $K_{ij}<0$.  
We also assume that the arrows always go from lower to higher node, namely
$K_{ij}>0$ for $i<j$, to avoid the oriented loops (superpotentials).
This system represents the bound states of $n$ distinguishable particles,
and will give basic building blocks of the Manschot-Pioline-Sen (MPS) formula \cite{Manschot:2010qz}.

The gauge group of the model is $G=\prod_{r=1}^n U(1)_r$
and there are $n$ integral variables $\phi_i$. Except for the center of mass, there are $n-1$ independent relative variables
and those are fixed by demanding $n-1$ independent BRST fixed point equations
\be
\phi_i -\phi_j +i\e^{ij}_{a_{ij}} = \frac{2\pi i}{\beta} n_{ij}, \quad \text{for }i<j\text{ and }n_{ij}\in\Z,
\ee
or equivalently 
\be
u_i u_j^{-1} x^{ij}_{a_{ij}} =1, \quad \text{for }i<j,
\ee
where $n-1$ combinations of two nodes are chosen.
A set of the indices $(i,j)$ which appear in the above $(n-1)$ equations 
is denoted by $I$.

From the BRST fixed point equations, we find that $q^{ij}_{a_{ij}}$ can get non-zero vev in the Higgs branch,
where $q^{ij}_{a_{ij}}$ is a scalar component of the chiral multiplet, corresponding to
the $a_{ij}$-th arrow between the $i$-th and $j$-th node.
The D-term equations with a constraint $\theta({\bf N})=\sum_{i=1}^n \zeta_i=0$ are
\be
\begin{split}
\sum_{j=2}^n |q^{1j}_{a_{1j}}|^2 &=\zeta_1,\\
\sum_{j=3}^n |q^{2j}_{a_{2j}}|^2-|q^{12}_{a_{12}}|^2 &=\zeta_2,\\
\vdots\\
-\sum_{i=1}^{n-1} |q^{in}_{a_{in}}|^2 &=\zeta_n,\\
\end{split}
\label{n-node D-term}
\ee
where $|q^{ij}_{a_{ij}}|=0$ if $(i,j)\not\in I$.
Thus, each fixed point is labeled by a possible set of $I$
if there exists a solution to the D-term equations (\ref{n-node D-term}).

Once the fixed point $I$ is found, we obtain the $T$-character 
\be
{\cal T}^K_n(\Phi^*(I)) = (t-1)
\left(
\sum_{(i,j) \in I}\sum_{l\neq a_{ij}}\frac{x^{ij}_l}{x^{ij}_{a_{ij}}}
+\sum_{(i,j) \not\in I} \sum_{l=1}^{K_{ij}}\frac{x^{ij}_l}{p_{ij}}
\right),
\ee
where 
$p_{ij}$ is a monomial of $x^{ij}_{a_{ij}}$'s and satisfies
$u_i u_j^{-1}p_{ij}=1$ for $(i,j)\not\in I$. 

Therefore, a contribution to the partition function at the fixed point is given by
\be
\Pcal^K_n(\Phi^*(I)) =
\prod_{(i,j)\in I}\prod_{l\neq a_{ij}}\frac{x^{ij}_{a_{ij}}-tx^{ij}_l}{x^{ij}_{a_{ij}}-x^{ij}_l}
\prod_{(i,j) \not\in I} \prod_{l=1}^{K_{ij}}\frac{p_{ij}-tx^{ij}_l}{p_{ij}-x^{ij}_l}.
\ee
We finally obtain the partition function
\be
\begin{split}
\hat{\Zcal}^K_n &= \sum_{I\in \text{fixed points}} \Pcal^K_n(\Phi^*(I))\\
&= \sum_{I\in \text{fixed points}}
\prod_{(i,j)\in I}\prod_{l\neq a_{ij}}\frac{x^{ij}_{a_{ij}}-tx^{ij}_l}{x^{ij}_{a_{ij}}-x^{ij}_l}
\prod_{(i,j) \not\in I} \prod_{l=1}^{K_{ij}}\frac{p_{ij}-tx^{ij}_l}{p_{ij}-x^{ij}_l}.
\end{split}
\ee

We check the above formula for a simple example of three nodes ($n=3$).
Possible sets of indices are $I_1=\{(1,2),(2,3)\}$, $I_2=\{(1,2),(1,3)\}$ and $I_3=\{(2,3),(1,3)\}$.
For $I_1=\{(1,2),(2,3)\}$, we require
\be
u_1u_2^{-1}x^{12}_{a_{12}}=1,\quad
u_2u_3^{-1}x^{23}_{a_{23}}=1,\quad
u_1u_3^{-1}p_{13}=1,
\ee
where $p_{13}=x^{12}_{a_{12}}x^{23}_{a_{23}}$.
The D-term equations become
\be
\begin{split}
|q^{12}_{a_{12}}|^2 &=\zeta_1,\\
|q^{23}_{a_{23}}|^2-|q^{12}_{a_{12}}|^2 &=\zeta_2,\\
- |q^{23}_{a_{23}}|^2 &=\zeta_3.\\
\end{split}
\ee
This has a solution only if $\zeta_1>0$ and $\zeta_3<0$.

Similarly, for $I_2=\{(1,2),(1,3)\}$ ($I_3=\{(2,3),(1,3)\}$),
there is a solution if $\zeta_1>0$, $\zeta_2<0$ and $\zeta_3<0$
($\zeta_1>0$, $\zeta_2>0$ and $\zeta_3<0$).
Thus, in a chamber of $\zeta_2>0$ ($\zeta_2<0$),
$I_1$ and $I_3$ ($I_1$ and $I_2$) are chosen for the fixed points.

The partition function  is evaluated by
\be
\begin{split}
\hat{\Zcal}_3^K&=\sum_{a_{12},a_{23}}
\prod_{l\neq a_{12}}\frac{x^{12}_{a_{12}}-tx^{12}_l}{x^{12}_{a_{12}}-x^{12}_l}
\prod_{l\neq a_{23}}\frac{x^{23}_{a_{23}}-tx^{23}_l}{x^{23}_{a_{23}}-x^{23}_l}
\prod_{l=1}^{K_{13}}\frac{x^{12}_{a_{12}}x^{23}_{a_{23}}-tx^{13}_l}{x^{12}_{a_{12}}x^{23}_{a_{23}}-x^{13}_l}\\
&\quad +\sum_{a_{23},a_{13}}
\prod_{l\neq a_{23}}\frac{x^{23}_{a_{23}}-tx^{23}_l}{x^{23}_{a_{23}}-x^{23}_l}
\prod_{l\neq a_{13}}\frac{x^{13}_{a_{13}}-tx^{13}_l}{x^{13}_{a_{13}}-x^{13}_l}
\prod_{l=1}^{K_{12}}\frac{x^{13}_{a_{13}}-tx^{12}_lx^{23}_{a_{23}}}{x^{13}_{a_{13}}-x^{12}_lx^{23}_{a_{23}}} \\
&=\frac{(1-t^{K_{23}})(1-t^{K_{12}+K_{13}})}{(1-t)^2},
\end{split}
\ee
for $\zeta_2>0$, and
\be
\hat{\Zcal}^K_3 = \frac{(1-t^{K_{12}})(1-t^{K_{23}+K_{13}})}{(1-t)^2},
\ee
for $\zeta_2<0$.
Here, all $x^{ij}_{a_{ij}}$ ($\e_{a_{ij}}^{ij}$) dependences disappear again. Thus, from (\ref{eq:originalp}), the original partition function becomes
\begin{align}
\Zcal_3^K&=t^{-\frac{1}{2}(K_{12}+K_{13}+K_{23}-2)}\hat{\Zcal}_3^K \notag \\
&=
\begin{cases}
\dfrac{(t^{K_{23}/2}-t^{-K_{23}/2})(t^{(K_{12}+K_{13})/2}-t^{-(K_{12}+K_{13})/2)})}{t^{1/2}-t^{-1/2}},\quad \text{for }\zeta_2 >0, \\[2ex]
\dfrac{(t^{K_{12}/2}-t^{-K_{12}/2})(t^{(K_{13}+K_{23})/2}-t^{-(K_{13}+K_{23})/2)})}{t^{1/2}-t^{-1/2}},\quad \text{for }\zeta_2 <0.
\end{cases}
\end{align}
Our result perfectly agrees with the Poincar\'{e} polynomial of the Higgs branch moduli \cite{Reineke}, or $g_\text{ref}(\alpha_1,\alpha_2,\alpha_3,y)$
in \cite{Manschot:2010qz} by setting $t^{1/2}=-y$.

At the end of this section, we give an expression of the Wilson loop operator vev.
The center of mass free Wilson loop operators are given by a polynomial of $u_iu_j^{-1}$
associated with each arrow (link). At the fixed point, $u_iu_j^{-1}=1/{x_{a_{ij}}^{ij}}$
for $(i,j)\in I$ and $u_i u_j^{-1} = 1/p_{ij}$ for $(i,j)\not\in I$.
If the polynomial evaluated at the fixed points $I$ is denoted by $P_I({x_{a_{ij}}^{ij}},p_{ij})$,
the vev of the Wilson loop is given by
\be
\left\langle
W
\right\rangle
=  \frac{1}{\hat{\Zcal}^K_n}\sum_{I\in \text{fixed points}}
\prod_{(i,j)\in I}\prod_{l\neq a_{ij}}\frac{x^{ij}_{a_{ij}}-tx^{ij}_l}{x^{ij}_{a_{ij}}-x^{ij}_l}
\prod_{(i,j) \not\in I} \prod_{l=1}^{K_{ij}}\frac{p_{ij}-tx^{ij}_l}{p_{ij}-x^{ij}_l}P_I({x_{a_{ij}}^{ij}},p_{ij}).
\ee
The expression is rather complicated, but can be evaluated exactly, in principle.

\section{Non-Abelian nodes}

In this section, we consider the quiver quantum mechanics including the non-Abelian nodes.

\subsection{Abelian and non-Abelian nodes: the Hall halo} \label{sec:abeliannonabelian}

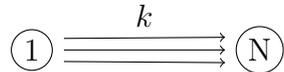
\begin{figure}[ht]
 \begin{center}
\begin{tikzpicture}
  \node (1) {$\circled{1}$};
  \node (N) [right of=1] {$\circled{N}$};
  \draw[->] (1.20) to node {$k$} (N.160);
  \draw[->] (1) to node {} (N);
  \draw[->] (1.340) to node {} (N.200);
\end{tikzpicture}
 \end{center}
\caption{$U(1)_1\times U(N)_2$ with $k$ arrows.}
\label{u1-un}
\end{figure}

We firstly consider the quiver quantum mechanics which is represented by two nodes of $U(1)_1\times U(N)_2$ with $k$ arrows in Fig.~\ref{u1-un}.
This case  can be regarded as a charge $k$ magnetic monopole
surrounded by $N$ mutually non-interacting electrons with charge one
in the Coulomb picture \cite{Denef:2002ru}.

If we first assume that $Z_2$ and $\bar{Z}_2$ are diagonal, then the BRST fixed point equations (\ref{BRST vector})
say that $Z_2=\bar{Z}_2=0$ because of the periodicity $Z_2(\tau+\beta)=Z_2(\tau)$
and $\bar{Z}_2(\tau+\beta)=\bar{Z}_2(\tau)$.
Then we can repeat the similar arguments in the Abelian case, and
we find the fixed point equations
\be
\begin{split}
\phi_1 - \phi_{2,1} +i\e_{a_1} &= \frac{2\pi i}{\beta} n_1,\\
\phi_1 - \phi_{2,2} +i\e_{a_2} &= \frac{2\pi i}{\beta} n_2,\\
\qquad \vdots\\
\phi_1 - \phi_{2,N} +i\e_{a_N} &= \frac{2\pi i}{\beta} n_N,
\end{split}
\ee
where $n_i \in \Z$. The $N$ indices $a_i$ must be chosen from different integers within $k$. Otherwise, the partition function vanishes because of the numerator of the 1-loop determinant for the vector multiplet.
Then, the fixed points are classified by a set of the indices $(a_1,a_2,\ldots,a_N)$.
We can assume that $a_1<a_2<\cdots < a_N$ by using the Weyl permutations.
For this ordering, the total number of the fixed points is $\binom{k}{N}$.
These fixed points are denoted by $\Phi^*_{a_1a_2\cdots a_N}$.

Since $[Z_2,\bar{Z}_2]=0$ in this case, the D-term equations are expressed only by the Higgs vev
\be
\begin{split}
\sum_{i=1}^N |q^i_{a_i}|^2 &= \zeta_1,\\
-\begin{pmatrix}
|q^1_{a_1}|^2\\
& |q^2_{a_2}|^2 &&\\
&& \ddots & \\
&&& |q^N_{a_N}|^2
\end{pmatrix}
&=\zeta_2 {\bf 1}_N,
\end{split}
\ee
in terms of the $N$  absolute values of $q^i_{a_i}$, associated with the BRST fixed points.
Under the constraint of the FI parameter $\theta({\bf N})=\zeta_1+ N\zeta_2=0$,
the D-term equations have a solution $|q^1_{a_1}|=|q^2_{a_2}|=\cdots=|q^N_{a_N}|=\sqrt{\zeta_1/N}$
only if $\zeta\equiv\zeta_1=-N\zeta_2>0$, otherwise no fixed point.

Using $u_1 = e^{\beta \phi_1}$ and $u_{2,i} = e^{\beta \phi_{2,i}}$,
the BRST fixed point equations give
\be
u_1 u_{2,1}^{-1} x_{a_1} = u_1 u_{2,2}^{-1} x_{a_2} = \cdots = u_1 u_{2,N}^{-1} x_{a_N}=1.
\ee
Thus, we obtain
\be
V_1 = u_1, \quad V_2 = u_1\sum_{i=1}^N x_{a_i}.
\ee
The fixed point data gives the $T$-character 
\be
\begin{split}
\Tcal_{1,N}(\Phi^*_{a_1 a_2\cdots a_N})
&= (1-t)\left(
1+\sum_{i,j=1}^N x_{a_i}x_{a_j}^{-1}- \sum_{n=1}^k \sum_{i=1}^N  x_n x_{a_i}^{-1} -1
\right)\\
&=(1-t)\left(
\sum_{i \neq j} x_{a_i}x_{a_j}^{-1} 
-\sum_{i=1}^N \sum_{n \neq a_i } x_n x_{a_i}^{-1}
\right).
\end{split}
\ee
The corresponding determinant in the partition function at each fixed point is
\be
\Pcal^k_{1,N}(\Phi^*_{a_1 a_2\cdots a_N}) = \frac{1}{N!}
\prod_{i \neq j} 
\frac{x_{a_i}-x_{a_j}}{x_{a_i}-tx_{a_j}}
\prod_{i=1}^N\prod_{n\neq a_i}\frac{x_{a_i}-tx_n}{x_{a_i}-x_n}.
\ee

Investigating various values of $N$ and $k$, we find that the total relative partition function
is given by the $t$-binomial coefficient
\be
\begin{split}
\hat{\Zcal}^k_{1,N} &= N!\sum_{a_1<a_2<\cdots<a_N} \Pcal_{1,N}^k(\Phi^*_{a_1 a_2\cdots a_N})\\
&=\frac{\prod_{j=1}^k (1-t^j)}{\prod_{j=1}^N (1-t^j)\prod_{j=1}^{k-N}(1-t^j)},
\end{split}
\ee 
where the $N!$ factor comes from the Weyl permutations.
This is the Poincar\'e polynomial of the Grassmannian $Gr(N,k)$, that is
the moduli space of the Higgs branch.
This also agrees with \cite{Denef:2002ru}.
 
So far we have assumed that $Z_2=\bar{Z}_2=0$.
One may wonder whether there is fixed point for the case of  $[Z_2,\bar{Z}_2]\neq 0$ when $\zeta_v\neq 0$.
This case corresponds to the poles in (\ref{integral formula}) coming from the 1-loop determinant of
the vector multiplet.
If there is a contribution from these poles, the classification of the fixed points becomes more
complicated due to off-diagonal components of $Z_2$ and $\bar{Z}_2$.
The fixed points for the non-commutative $Z_2$ and $\bar{Z}_2$ are classified as similar as the
Nekrasov formula by using the Young diagrams\cite{Nekrasov:2002qd,Nekrasov:2003rj}, 
but
the partition function should vanish in this branch if these poles are rejected by the F-term and D-term conditions.

Let us now discuss the possibility of the branch with $[Z_2,\bar{Z}_2]\neq 0$ in our quiver model.
First of all, from the BRST equation, requiring the periodicity of $Z_2$ and $\bar{Z}_2$, we obtain
\begin{align}
[\Phi_2, Z_2]+i\e Z_2 &= 0,\\
[\Phi_2, \bar{Z}_2]-i\e \bar{Z}_2 &= 0.
\end{align}
The above equations have generally blockwise solutions up to the Weyl permutations
\begin{align}
&\Phi_2 =
\begin{pmatrix}
\alpha_1{\bf 1}_{n_1} + i\e L_0^{(n_1)} & & &\\
& \alpha_2{\bf 1}_{n_2}+i\e L_0^{(n_2)} & &\\
& & \ddots &\\
& & & \alpha_d{\bf 1}_{n_d}+i\e L_0^{(n_d)} 
\end{pmatrix},\\
&Z_2 =
\begin{pmatrix}
L_-^{(n_1)} & & &\\
& L_-^{(n_2)} & &\\
& & \ddots &\\
& & & L_-^{(n_d)} 
\end{pmatrix},
\quad
\bar{Z}_2 =
\begin{pmatrix}
L_+^{(n_1)} & & &\\
& L_+^{(n_2)} & &\\
& & \ddots &\\
& & & L_+^{(n_d)} 
\end{pmatrix},
\end{align}
where
$(L_0^{(n_s)})_{j,j}=j-1$ for $j=1,\cdots,n_s$, and $(L_-^{(n_s)})_{j+1,j}=z_{s,j}$ and
$(L_+^{(n_s)})_{j,j+1}=\bar{z}_{s,j}$ for $j=1,\cdots,n-1$, 
otherwise zero.
Here $n_s$'s satisfy $\sum_{s=1}^d n_s =N$, then the fixed points are classified in terms of the partitions
of $N$, namely the Young diagrams of $N$ boxes.

Secondly, we have to find solutions which satisfy the other BRST equations for the chiral multiplet, F-term and D-term conditions at the same time.
Solving the F-term condition
\be
Z_1q_a -q_a Z_2=0,
\ee
and BRST equations
\be
\del_\tau q_a + \phi_1 q_a - q_a \Phi_2+i\e_a q_a=0,
\ee
we find that $Z_1=0$ and $\alpha_{s} = \phi_1+i\e_{a_s}$ ($s=1,\cdots,d$),
and also each $q_{a_s}$ has a highest component only, namely $q_{a_s}=(h_{a_s},0,\ldots,0)$.

For the above solutions, however, there is no solution to the D-term equations. Indeed,
for each $n$-dimensional block, we see that
\be
\begin{split}
\mu^{(n)}_{\R,2}&=\frac{1}{2}[L_-^{(n)},L_+^{(n)}]
-g^2  \bar{q}_{a}q_{a}-\zeta_2\\
&=\begin{pmatrix}
-\frac{1}{2}|z_{1}|^2 - g^2|h_a|^2 -\zeta_2 & & &\\
&\frac{1}{2}( |z_1|^2-|z_2|^2) -\zeta_2&&\\
&& \ddots & \\
&&& \frac{1}{2}|z_n|^2 -\zeta_2
\end{pmatrix}=0,
\end{split}
\ee
which does not have any solution for both $\zeta_2>0$ and $\zeta_2<0$.
Thus we can conclude that there is no Nakajima-Nekrasov type fixed point in our quiver matrix model.
This is a consequence of the orthogonality between the Higgs branch ($Z=0$ and $q\neq0$)
and the Coulomb branch ($Z\neq 0$ and $q=0$) from the F-term condition.\footnote{
In the Nekrasov formula, there is no F-term condition $qZ=0$ like our quiver model. Then
we can choose lowest component of $q_a$ as a solution to the D-term equations.}

%
%
%

The Wilson loop for the non-Abelian part is obtained by a trace over a representation $R$.
In the diagonal gauge, it is written in terms of a symmetric polynomial of $u_{2,i}$,
that is, a Schur polynomial associated with the representation (Young diagram) $R$.
If we denote the polynomial by $s_R(u_{2,1},u_{2,1},\ldots,u_{2,N})$, the center of mass motion free Wilson loop
is expressed by $u_1^{-d}s_R(u_{2,1},u_{2,1},\ldots,u_{2,N})$, where $d$ is the degree of the polynomial.
At the fixed point, the Wilson loop becomes simply $s_R(x_{a_1},x_{a_2},\ldots,x_{a_N})$ in $N$ variables chosen
from $k$.
Therefore, we find the Wilson loop in the representation $R$ is evaluated by
\be
\begin{split}
\left\langle
W_R
\right\rangle
 &= \frac{1}{\hat{\Zcal}^k_{1,N}}\sum_{a_1<a_2<\cdots<a_N}\prod_{i \neq j} 
\frac{x_{a_i}-x_{a_j}}{x_{a_i}-tx_{a_j}}
\prod_{i=1}^N\prod_{n\neq a_i}\frac{x_{a_i}-tx_n}{x_{a_i}-x_n}
s_R(x_{a_1},x_{a_2},\ldots,x_{a_N}).
\end{split}
\ee
This results might be interesting from a point of the view of the physics and
the symmetric polynomial of $t$ and $x_{a_i}$'s,
but we do not pursue the Wilson loops anymore.
We will concentrate on the partition function only in the following.

\subsection{Coprime dimension vector} \label{sec:coprime}

We generalize our arguments to the non-Abelian nodes.
The classification of the fixed points in quiver theory only with non-Abelian nodes
is rather complicated. So we here explain only a few examples.

We first start with two nodes with a coprime dimension vector, that is
$U(2)_1\times U(3)_2$ with $k$ arrows. There are
five integral variables of $\phi_{1,i}~(i=1,2)$ and $\phi_{2,i'}~(i'=1,2,3)$. Except for the center of mass, 
we can give four fixed point equations to solve in $\phi_{1,i}$ and $\phi_{2,i'}$.
For the exponent variables, a possible set of the BRST fixed point equations is
\be
\begin{array}{ll}
u_1 \tilde{u}_1^{-1} x_a =1,&
u_1 \tilde{u}_2^{-1} x_b =1,\\
u_2 \tilde{u}_1^{-1} x_c =1,&
u_2 \tilde{u}_3^{-1} x_d =1.
\label{u2-u3 BRST fixed points}
\end{array}
\ee
We only consider the case of $a\neq b$, $a\neq c$, and $c\neq d$ because otherwise the partition function vanishes. 
The corresponding squarks $q^a_{11}$, $q^b_{12}$, $q^c_{21}$, and $q^d_{23}$ could have the 
vevs as a consequence of the D-term equations. The D-term equation becomes
\begin{align}
\begin{pmatrix}
|q^a_{11}|^2+|q^b_{12}|^2 & 0 \\
0 & |q^c_{21}|^2+ |q^d_{23}|^2
\end{pmatrix}
&=\zeta_1 {\bf 1}_2,\\
-\begin{pmatrix}
|q^a_{11}|^2+|q^c_{21}|^2 & 0 & 0\\
0 & |q^b_{12}|^2 & 0\\
0 & 0 & |q^d_{23}|^2
\end{pmatrix}
&=\zeta_2 {\bf 1}_3.
\end{align}
Under the FI parameter constraint $\theta({\bf N})=2\zeta_1+3\zeta_2=0$,
the D-term equations have a solution
$|q^a_{11}|=|q^c_{21}|=\sqrt{\zeta_1/3}$ and $|q^b_{12}|=|q^d_{23}|=\sqrt{2\zeta_1/3}$
only if $\zeta\equiv 2\zeta_1=-3\zeta_2>0$.

If $\zeta>0$, there exist $4!\binom{k}{4}+3\cdot 3!\binom{k}{3}+2!\binom{k}{2}=k(k-1)^3$
fixed points in total by choosing appropriate $a,b,c,d$ from integers within $k$.
The fixed point set of this type is denoted by $\Phi^*[^{ab}_{c~d}]$,
which reflects the structure of the BRST fixed points (\ref{u2-u3 BRST fixed points})
in a $2\times 3$ matrix of the labels.
We can also find that there are other 5 fixed point sets of the similar structure like
$\Phi^*[^{ab}_{~cd}]$, $\Phi^*[^{ab}_{~cd}]$,  $\Phi^*[^{a~b}_{~cd}]$, etc.
This classification of the fixed points is already known in the mathematical literature \cite{Weist}.
These fixed points are depicted in the diagram:
 
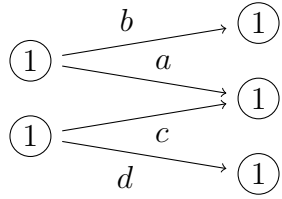
\begin{center}
\begin{tikzpicture}
  \node (11) at (0,0) {$\circled{1}$};
  \node (12) at (0,-1) {$\circled{1}$};
  \node (21) at (3,0.5) {$\circled{1}$};
  \node (22) at (3,-0.5) {$\circled{1}$};
  \node (23) at (3,-1.5) {$\circled{1}$};
  \draw[->] (11) to node {$b$} (21);
  \draw[->] (11) to node {$a$} (22);
  \draw[->] (12) to node [swap] {$c$} (22);
  \draw[->] (12) to node [swap] {$d$} (23);
\end{tikzpicture}
 \end{center}
All of the fixed points of this kind will give the same contribution to the $T$-character and partition function.
So we here evaluate the contribution from the fixed point set of $\Phi^*[^{ab}_{c~d}]$ only as follows.

Solving (\ref{u2-u3 BRST fixed points}), we find
\be
\begin{split}
V_1 &= \tilde{u}_3(x_a^{-1}x_cx_d^{-1}+x_d^{-1}),\\
V_2 &=\tilde{u}_3(x_cx_d^{-1}+x_a^{-1}x_bx_cx_d^{-1}+1),
\end{split}
\ee
where the remaining $\tilde{u}_3$ corresponds to the degree of the center of mass.
Using this solution, we obtain the $T$-character at the fixed point
\begin{multline}
\Tcal^k_{2,3}(\Phi^*[^{ab}_{c~d}])
=(1-t)\Bigg(
\frac{x_a x_d}{x_b x_c}+\frac{x_bx_c}{x_ax_d}
-\sum_{l\neq a,b,c} \frac{x_l}{x_a}-\sum_{l\neq a,b}\frac{x_l}{x_b}-\sum_{l\neq a,c,d}\frac{x_l}{x_c}
-\sum_{l\neq c,d}\frac{x_l}{x_d}\\
-\sum_{l=1}^k\frac{x_bx_l}{x_a x_c}-\sum_{l=1}^k\frac{x_c x_l}{x_b x_d}
\Bigg).
\end{multline}
Then the contribution to the partition function from this fixed point is given by
\be
\begin{split}
\Pcal^{k}_{2,3}(\Phi^*[^{ab}_{c~d}]) &=\frac{1}{2!3!}
\frac{x_ax_d-x_bx_c}{x_ax_d-tx_bx_c}\frac{x_bx_c-x_ax_d}{x_bx_c-tx_ax_d}\\
&\qquad \times
\prod_{l\neq a,b,c}\frac{x_a-t x_l}{x_a -x_l}
\prod_{l\neq a,b}\frac{x_b-t x_l}{x_b -x_l}
\prod_{l\neq a,c,d}\frac{x_c-t x_l}{x_c -x_l}
\prod_{l\neq c,d}\frac{x_d-t x_l}{x_d -x_l}\\
&\qquad \times
\prod_{l=1}^k\frac{x_ax_c-tx_bx_l}{x_ax_c-x_bx_l}
\prod_{l=1}^k\frac{x_bx_d-tx_cx_l}{x_bx_d-x_cx_l}.
\end{split}
\ee

In this quiver gauge theory, it is important to notice that there is yet an another kind of the fixed points.
To find it, we go back to the original expression of the partition function,
\be
\begin{split}
\Zcal_{2,3}^k&= \frac{1}{2!3!}\int
\prod_{i=1}^2 \frac{d\phi_{1,i}}{2\pi i}
\prod_{i'=1}^3 \frac{d\phi_{2,i'}}{2\pi i'}
 \prod_{i\neq j}^2\frac{\sinh \frac{\beta}{2}(\phi_{1,i}-\phi_{1,j})}{\sinh\frac{\beta}{2}(\phi_{1,i}-\phi_{1,j}+i\e)}\prod_{i'\neq j'}^3\frac{\sinh \frac{\beta}{2}(\phi_{2,i'}-\phi_{2,j'})}{\sinh\frac{\beta}{2}(\phi_{2,i'}-\phi_{2,j'}+i\e)} \\
&\qquad \times \prod_{a=1}^k\prod_{i=1}^2\prod_{i'=1}^3\frac{\sinh\frac{\beta}{2}(\phi_{1,i}-\phi_{2,i'}+i(\e +\e_a))}{\sinh\frac{\beta}{2}(\phi_{1,i}-\phi_{2,i'}+i\e_a)}.
\end{split}
\ee
At first, if we pick up  the following three poles
\be
\begin{split}
\phi_{1,1}-\phi_{2,1}+i\e_a&=0, \\
\phi_{1,1}-\phi_{2,2}+i\e_b&=0, \\
\phi_{1,1}-\phi_{2,3}+i\e_c&=0,
\end{split}
\ee
where $a\neq b \neq c$, then we find a factor of the residue
\begin{align}
\frac{\sinh^2(\phi_{1,1}-\phi_{1,2})}{\sinh^3(\phi_{1,1}-\phi_{1,2})}=\frac{1}{\sinh(\phi_{1,1}-\phi_{1,2})},
\end{align}
which gives a new pole
\begin{align}
 \phi_{1,1}-\phi_{1,2}=0.
\end{align}

In terms of the exponent variables, this type of the  fixed point set  is given by
four equations
\be
u_1=u_2,\quad
u_1\tilde{u}_1^{-1}x_a=1,\quad
u_1\tilde{u}_2^{-1}x_b=1,\quad
u_1\tilde{u}_3^{-1}x_c=1.
\label{abc fixed points}
\ee
In this phase, the gauge symmetry of the $U(2)$ factor degenerates due to the fixed point equation $u_1=u_2$
and the D-term equations are modified to
\begin{align}
|q^a_{11}|^2 + |q^b_{12}|^2 + |q^c_{13}|^2&= 2\zeta_1,\\
-\begin{pmatrix}
|q^a_{11}|^2 & &\\
&  |q^b_{12}|^2 & \\
& & |q^c_{13}|^2
\end{pmatrix}
&=\zeta_2.
\end{align}
This has a solution $|q^a_{11}| = |q^b_{12}| = |q^c_{13}| = \sqrt{\zeta/3}$
only if $\zeta>0$.
We denote $\Phi^*[abc]$ as this kind of the fixed points.
We also have the similar fixed points by exchanging the role of $u_1$ and $u_2$.
The total number of these fixed points is $2\cdot 3! \binom{k}{3}$ and this also classified in \cite{Weist}
by the diagram:
 
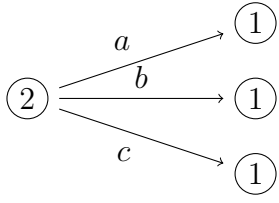
\begin{center}
\begin{tikzpicture}
  \node (2) at (0,0) {$\circled{2}$};
  \node (21) at (3,1) {$\circled{1}$};
  \node (22) at (3,0) {$\circled{1}$};
  \node (23) at (3,-1) {$\circled{1}$};
  \draw[->] (2) to node {$a$} (21);
  \draw[->] (2) to node {$b$} (22);
  \draw[->] (2) to node [swap] {$c$} (23);
\end{tikzpicture}
 \end{center}

At the fixed points $\Phi^*[abc]$, solving the BRST fixed point equation (\ref{abc fixed points}),
we find
\begin{align}
V_1 &= 2 \tilde{u}_3 x_c^{-1}, \\
V_2 &= \tilde{u}_3  ( x_a x_c^{-1} + x_b x_c^{-1}+1),
\end{align}
which give the $T$-character
\begin{multline}
{\Tcal}^k_{2,3}(\Phi^*[abc])\\ 
=(1-t)
\left(\frac{x_a}{x_b}+\frac{x_b}{x_a}+\frac{x_b}{x_c}+\frac{x_c}{x_b}+\frac{x_c}{x_a}+\frac{x_a}{x_c}
-2\sum_{l\neq a}\frac{x_l}{x_a}-2\sum_{l\neq b}\frac{x_l}{x_b}-2\sum_{l\neq c}\frac{x_l}{x_c}\right).
\end{multline}
Then we have
\be
\begin{split}
\Pcal^k_{2,3}(\Phi^*[abc]) &= \frac{1}{2!3!}\frac{x_a-x_b}{x_a-tx_b}\frac{x_b-x_a}{x_b-tx_a}\frac{x_b-x_c}{x_b-tx_c}\frac{x_c-x_b}{x_c-tx_b}
\frac{x_c-x_a}{x_c-tx_a}\frac{x_a-x_c}{x_a-tx_c}\\
&\qquad \times
\prod_{l\neq a}\frac{(x_a-tx_l)^2}{(x_a-x_l)^2}
\prod_{l\neq b}\frac{(x_b-tx_l)^2}{(x_b-x_l)^2}
\prod_{l\neq c}\frac{(x_c-tx_l)^2}{(x_c-x_l)^2}.
\end{split}
\ee

The classification of the fixed points for this quiver theory is finished. Thus we obtain the total relative
partition function by a summation over $6k(k-1)^3+2k(k-1)(k-2)=2k(k-1)(3k^2-5k+1)$ total fixed points
\be
\hat{\Zcal}^k_{2,3} = 6\sum_{a\neq b,a\neq c,c\neq d} \Pcal^{k}_{2,3}(\Phi^*[^{ab}_{c~d}])
+2\sum_{\substack{a\neq b,b\neq c,c\neq a}}\Pcal^k_{2,3}(\Phi^*[abc]),
\ee
which gives the Poincar\'e polynomial of the Higgs branch moduli space.
In the $t\to 1$ ($\e\to 0$) limit, $\Pcal^k_{2,3}$ at each fixed point contributes by $+1$ to the partition function.
So the index of the quiver theory is given by the total number of the fixed points divided
by the order of the Weyl group
\be
\lim_{t\to 1} \hat{\Zcal}^k_{2,3} = \frac{1}{6}k(k-1)(3k^2-5k+1).
\ee
This counting agrees with the Euler characteristic derived in \cite{Reineke, Weist}.

We finally give some explicit results of the partition function for smaller $k$ in the following:
\begin{align}
\Zcal^1_{2,3} &= 0, \\
\Zcal^2_{2,3} &= 1,\\
\Zcal^3_{2,3} &= \frac{1}{t^3}(1+t+3t^2+3t^3+3t^4+t^5+t^6),\\
\Zcal^4_{2,3} &= \frac{1}{t^6}(1+t+3 t^{2}+4 t^3+7 t^4+8 t^5+10 t^6+8 t^7+7 t^8+4 t^9+3 t^{10}+t^{11}+t^{12}),\\
\Zcal^5_{2,3} &= \frac{1}{t^9}(1+t+3 t^2+4 t^3+7 t^4+9 t^5+14 t^6+16 t^7+20 t^8+20 t^9\notag \\
&\qquad\qquad\qquad+20 t^{10}+16 t^{11}+14 t^{12}+9 t^{13}+7 t^{14}+4 t^{15}+3 t^{16}+t^{17}+t^{18}).
\end{align}
These results  agree with Reineke's formula \cite{Reineke,Denef:2002ru} and the wall crossing formal{\ae} (\ref{eq:app231})-(\ref{eq:app235}) in Appendix \ref{sec:app},
  by setting $t^{1/2}=-y$ \cite{Kontsevich:2008fj, Joyce:2008pc, Manschot:2011xc}.

\subsection{Non-coprime dimension vector} \label{sec:noncoprime}

Next we treat the case of ${\bf N}=(2,2)$, which has a common divisor among the dimensions.
The first type of the BRST fixed points comes from poles of the chiral multiplet
\be
\begin{array}{ll}
u_1 \tilde{u}_1^{-1} x_a =1,&
u_1 \tilde{u}_2^{-1} x_b =1,\\
u_2 \tilde{u}_1^{-1} x_c =1,& 
\end{array} \label{chiralfix}
\ee
which can be solved by
\begin{align}
V_1 &= \tilde{u}_2 (x_b^{-1} + x_a x_b^{-1}x_c^{-1}), \\
V_2 &= \tilde{u}_2 (x_a x_b^{-1}+1).
\end{align}
We only consider the case of $a\neq b$ and $a\neq c$ because otherwise the partition function vanishes. The D-term constraint becomes
\begin{align}
\begin{pmatrix}
|q_{11}^a|^2+|q_{12}^b|^2 & 0 \\
0 & |q_{21}^c|^2
\end{pmatrix}
&=
\begin{pmatrix}
\zeta_1 & 0 \\
0 & \zeta_1
\end{pmatrix}, \\
-
\begin{pmatrix}
|q_{11}^a|^2+|q_{21}^c|^2 & 0 \\
0 & |q_{12}^b|^2
\end{pmatrix}
&=
\begin{pmatrix}
\zeta_2& 0 \\
0 & \zeta_2
\end{pmatrix},
\end{align}
where $\theta ({\bf N})=2\zeta_1+2\zeta_2=0$. Solving these equations, we find $|q_{11}^a|^2=0,~|q_{12}^b|^2=|q_{21}^c|^2=\zeta$, where $\zeta\equiv \zeta_1=-\zeta_2$,
but we have to resolve the singular solution $|q_{11}^a|^2=0$ because the FI parameters are on the wall of marginal stability \cite{Hwang:2014uwa}. To do this,
we slightly modify the FI parameters as follows:
\begin{align}
\zeta_1 \mathbf{1}_2 &\to 
\begin{pmatrix}
\zeta+\delta & 0 \\
0 & \zeta -\delta
\end{pmatrix}, \label{eq: FI1} 
\end{align}
where $\delta > 0$. The detail of the above choice of the FI parameters is explained in the Appendix \ref{mod}.
Note that this modification does not spoil the condition $\theta ({\bf N})=0$ and 
this infinitesimal parameter $\delta$  corresponds to the concept of the $\theta$-stability. Then, the solution of the D-term constraint becomes
\begin{align}
|q_{11}^a|^2&=\delta, \label{q11afin} \\
|q_{12}^b|^2&=\zeta , \\
|q_{21}^c|^2&=\zeta -\delta.
\end{align}
Thus, the vev of the squarks $q^a_{11}$, $q^b_{12}$, and $q^c_{21}$
satisfies the D-term equation if  $\zeta > \delta>0$.
The above solution is valid for the localization fixed point, which 
is denoted by $\Phi^*[^{ab}_c]$ and depicted by a diagram:
 
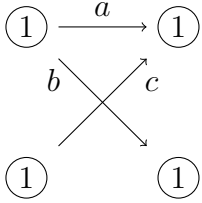
\begin{center}
\begin{tikzpicture}
  \node (11) at (0,0) {$\circled{1}$};
  \node (12) at (0,-2) {$\circled{1}$};
  \node (21) at (2,0) {$\circled{1}$};
  \node (22) at (2,-2) {$\circled{1}$};
  \draw[->] (11) to node {$a$} (21);
  \draw[->] (11) to node [above left] {$b\quad $} (22);
  \draw[->] (12) to node [above right] {$\quad c$} (21);
\end{tikzpicture}
 \end{center}
 There are $k(k-1)^2$ fixed points of this kind in total.

The $T$-character at this fixed point reduces to
\be
{\cal T}_{2,2}^k (\Phi^*[^{ab}_c])
=(t-1)\left(
\sum_{l\neq a,b,c}\frac{x_l}{x_a}
+\sum_{l\neq a,b}\frac{x_l}{x_b}
+\sum_{l\neq a,c}\frac{x_l}{x_c}
+\sum_{l=1}^k\frac{x_a x_l}{x_b x_c}\right).
\ee
So the 1-loop determinant at the fixed point is
\be
\Pcal^k_{2,2}(\Phi^*[^{ab}_c])
=\frac{1}{2!2!}
\prod_{l\neq a,b,c}\frac{x_a-t x_l}{x_a-x_l}
\prod_{l\neq a,b}\frac{x_b-t x_l}{x_b-x_l}
\prod_{l\neq a,c}\frac{x_c-t x_l}{x_c-x_l}
\prod_{l= 1}^k\frac{x_bx_c-t x_ax_l}{x_bx_c-x_ax_l},
\ee
which contributes to the partition function.

There seems to be three other fixed point sets such as $\Phi^*[^{ab}_{~c}]$, $\Phi^*[_{ab}^{c}]$
and $\Phi^*[_{ab}^{~c}]$, which give the same contribution to the character and partition function,
because of the Weyl permutations. However, after the modification of the FI parameters, we find that only $\Phi^*[^{ab}_{~c}]$ satisfies the D-term constraint.\footnote{The fixed point $\Phi^*[_{bc}^{a}]$ does not satisfy the D-term constraint because  one finds $|q_{21}^b|^2=-\delta<0$. Also, the fixed point $\Phi^*[_{bc}^{~a}]$ is rejected because $|q_{22}^b|^2=-\delta<0$. }
Therefore, there are two times the contributions of $\Pcal^k_{2,2}(\Phi^*[^{ab}_c])$ to the partition function in total.

Another kind of the fixed point is given by
\be
u_1 u_2^{-1} t =1, \quad
u_1\tilde{u}_1^{-1} x_a=1, \quad
u_2 \tilde{u}_2^{-1} x_a =1,
\label{u2u2 BRST}
\ee
which are solved by
\begin{align}
V_1&= \tilde{u}_2 (t^{-1}+1)x_a^{-1},
\label{u2u2 solution 1}\\
V_2&= \tilde{u}_2 (t^{-1}+1).
\label{u2u2 solution 2}
\end{align}
The first equation in (\ref{u2u2 BRST}) and solution (\ref{u2u2 solution 1})
and  (\ref{u2u2 solution 2}) means no longer $Z_1=Z_2=0$, but
they contains off-diagonal elements
\be
Z_1 = \begin{pmatrix}
0 & z^1_{12}\\
0 &0
\end{pmatrix},\quad
Z_2 = \begin{pmatrix}
0 & z^2_{12}\\
0 &0
\end{pmatrix}.
\ee
The F-term equation $Z_1 q_a  - q_a Z_2=0$ is satisfied if 
\begin{align}
q^a_{22}z^1_{12} - q^a_{11}z^2_{12}=0. \label{22Fterm}
\end{align}
And the D-term equation becomes
\be
\begin{split}
\frac{1}{2g^2}|z^1_{12}|^2 + |q^a_{11}|^2 &= \zeta+\delta,\\
-\frac{1}{2g^2}|z^1_{12}|^2 + |q^a_{22}|^2 &= \zeta -\delta,\\
\frac{1}{2g^2}|z^2_{12}|^2 - |q^a_{11}|^2 &= -\zeta,\\
-\frac{1}{2g^2}|z^2_{12}|^2 - |q^a_{22}|^2 &= -\zeta ,
\end{split}
\ee
where we have used the modified FI parameters (\ref{eq: FI1}). Solving the F-term and D-term constraints, we find
\begin{align}
|q_{11}^a|^2&=\frac{2\zeta(\zeta+\delta)}{2\zeta +\delta},~~~~~~~~~~|q_{22}^a|^2=\frac{2\zeta^2}{2\zeta+\delta}, \\
|z^1_{12}|^2&=2g^2\frac{\delta(\zeta+\delta)}{2\zeta+\delta},~~~~~~~~|z^2_{12}|^2=2g^2\frac{\zeta \delta}{2\zeta+\delta} .
\end{align}
Therefore, if $\zeta>0$, this fixed point satisfies the constraints.

%

In this way, we obtain the extra $k$ fixed points, which are denoted by $\Phi^*[^{a~}_{~a}]$ and a diagram:
 \begin{center}
\begin{tikzpicture}
  \node (1) at (0,0) {$\circled{2}$};
  \node (2) at (2,0) {$\circled{2}$};
  \draw[->] (1) to node {$a$} (2);
\end{tikzpicture}
 \end{center}
In the case of the coprime dimension vector, the Nakajima-Nekrasov type fixed points $Z_v\neq 0$
are rejected by the F-term and D-term equations. But we should
take into account the Nakajima-Nekrasov type fixed points in the present non-coprime case.
Using the solution (\ref{u2u2 solution 1})
and  (\ref{u2u2 solution 2}), we find the $T$-character
\be
\begin{split}
{\cal T}_{2,2}^k (\Phi^*[^{a~}_{~a}])
=t^{-1}-t^2
-(t^{-1}+1-t-t^2)\sum_{l=1}^k\frac{x_l}{x_a},
\end{split}
\ee
and the 1-loop determinant at the fixed point
\be
\begin{split}
\Pcal^k_{2,2}(\Phi^*[^{a~}_{~a}])
&=-\frac{1}{2!2!}
\frac{1}{t(1+t)}
\prod_{l\neq a}\frac{(x_a-t x_l)(x_a-t^2 x_l)}{(x_a-t^{-1}x_l)(x_a-x_l)}.
\end{split}
\ee
Note here that the contribution to the index from each fixed point is $-\frac{1}{2}$ up to the Weyl factor.

There seems to be three other fixed point sets of the same type such as
\begin{eqnarray}
&\bullet & u_1u_2^{-1}t^{-1}=1,\quad u_1\tilde{u}_1^{-1}x_a=1, \quad u_2\tilde{u}_2^{-1}x_a=1, \label{vectorfail1} \\
&\bullet & u_1u_2^{-1}t=1,\quad u_1\tilde{u}_2^{-1}x_a=1, \quad u_2\tilde{u}_1^{-1}x_a=1, \label{eq:vectoradd} \\
&\bullet & u_1u_2^{-1}t^{-1}=1,\quad u_1\tilde{u}_2^{-1}x_a=1, \quad u_2\tilde{u}_1^{-1}x_a=1.  \label{vectorfail2}
\end{eqnarray}
However, under the modification of the FI parameters, we find that only (\ref{eq:vectoradd}) satisfies the constraints (see Appendix \ref{mod}),
and it gives the same contribution as the fixed point $\Phi^*[^{a~}_{~a}]$ (twice of $\Pcal^k_{2,2}(\Phi^*[^{a~}_{~a}])$ in total).

Combining all of contributions from the fixed points, we obtain the total partition function
\be
\hat{\Zcal}^k_{2,2} = 2\sum_{a\neq b,a\neq c}\Pcal^k_{2,2}(\Phi^*[^{ab}_c])+2\sum_a \Pcal^k_{2,2}(\Phi^*[^{a~}_{~a}]).
\label{u2u2 partition function}
\ee
We again can see all $x_a$ dependences disappear in the final results.
As mentioned above, in the limit of $t\to 1$, the former term in (\ref{u2u2 partition function}) contributes to the index by $+1$,
while the latter term contributes by $-\frac{1}{2}$. Thus we find the index (Euler characteristic) for this model
\be
\lim_{t\to 1}\hat{\Zcal}^k_{2,2} = \frac{1}{4}k(2k^2-4k+1).
\ee

We finally give our results of the partition function for smaller $k$:
\begin{align}
\Zcal^1_{2,2} &= -\frac{t^{1/2}}{2(1+t)},\\
\Zcal^2_{2,2} &= \frac{t^{-1/2}}{2(1+t)}\left( 1+t^2 \right),\\
\Zcal^3_{2,2} &= \frac{t^{-5/2}}{2(1+t)}\left(1+t^2+t^4\right) \left(2 +3 t+2 t^2\right),\\
\Zcal^4_{2,2}&=\frac{t^{-9/2}}{2(1+t)}(1+t^2+t^4+t^6)(2+4t+5t^2+4t^3+2t^4), \\
\Zcal^5_{2,2} &= \frac{t^{-13/2}}{2(1+t)}\left(1+t^2+t^4+t^6+t^8\right)
\left(2 +4 t+6 t^2+7 t^3+6 t^4+4 t^5+2t^6\right).
\end{align}
In this case, we cannot directly compare our results with Reineke's formula because Reineke's formula
gives the so-called stack invariant in the case of
the non-coprime dimension vector \cite{Reineke}.
We must compare our result with
the rational refined index argued
in \cite{Moore:1998et,Manschot:2010qz}.
(See Appendix \ref{sec:app}.)
In fact, our results perfectly
agree with the special case of the wall crossing formul{\ae}
(\ref{eq:app221})-(\ref{eq:app225}) by setting $t^{1/2}=-y$. 

\section{Conclusion and Discussion}

In this paper, we have derived the exact partition functions and  expectation values of the Wilson loop operators of the quiver quantum mechanics
in the Higgs phase by the localization techniques. We have considered several examples of the quiver quantum mechanics which include only Abelian nodes in section \ref{sec:abelian}, a non-Abelian and an Abelian node in section  \ref{sec:abeliannonabelian}, and only non-Abelian nodes in section \ref{sec:coprime} and \ref{sec:noncoprime}.
We have found that those partition functions give the (refined) Witten index of the quiver theory physically,
or Poincar\'e polynomial of the quiver moduli mathematically.
Our results also agree with ones from the wall crossing formula.

To regularize the path integral of the matrix quantum mechanics, we have gauged the R-symmetries for both
vector multiplet and chiral multiplet. The background gauge fields of the R-symmetries
lift up the flat direction and the localization fixed points become isolated.
In terms of the residue integral, this means that all poles are decomposed into the simple poles.
We have considered the chiral multiplets with the general R-charges, but the all final results
are independent of these R-charges, except for the R-charge of the vector multiplet,
and thus its background gauge field becomes the refined parameter of the index.
This disappearance of the R-charges from the index is a significant property in our results although the vevs of Wilson loops depend on those R-charges.
As we have seen in the explicit examples, it is very subtle task to scrutinize the D-term
and F-term conditions in the non-Abelian cases. However, if we fail to take the correct fixed points into account,
the R-charge dependence never disappears. Thus, we can use this property for a criterion
of correctness of the partition functions.  

We have considered the localization in the Higgs branch by solving directly the linear (quiver) sigma model.
On the other hand, the quiver theory goes to the Coulomb phase in the IR limit $g_s\gg 1$.
The effective theory in the Coulomb phase can be written in terms of a non-linear sigma model
of the Abelian vector multiplets, that describe more directly charged BPS particles in ${\cal N}=2$ supergravity.
The localization in the Coulomb branch has been discussed already in \cite{deBoer:2008zn,Manschot:2010qz},
but we would like to revisit the Coulomb branch localization explicitly from a physical (matrix model)
point of view,
since we believe that  it is a key to understand the correspondence between the gauge theory
and gravity via the localization.
Deeper understandings of the correspondence may make clear how to emerge the supergravity
or M-theory from the matrix quantum mechanics, proposed in \cite{Banks:1996vh,Berenstein:2002jq}.

\section*{Acknowledgements}

We would like to thank D.~Gang, S.~Matsuura, T.~Misumi, T.~Nishinaka, N.~Sakai, S.~Yamaguchi, P.~Yi, D.~Yokoyama and Y.~Yoshida for useful discussions and comments.
This work of KO was supported in part by JSPS KAKENHI Grant Number 14485514.

\section*{Appendix}

\appendix

\section{Wall crossing formul\ae} \label{sec:app}
Let us denote an index of  BPS states as $\Omega(\gamma ; t^a)$, where $\gamma$ is the charge vector of the BPS states and $t^a$ are some parameters such as scalar moduli at spatial infinity. When the parameters $t^a$ come across the walls of marginal stability, which are codimension one subspaces of the parameter space, some of the BPS states decay and the index $\Omega(\gamma ; t^a)$ jumps. The formula for the change of the index is called the wall crossing formula \cite{Denef:2007vg}. This formula can be generalized  for the refined index $\Omega_{\mathrm{ref}}(\gamma,y)$, which keeps track of the angular momentum of the BPS states.
The generic wall crossing formula for the (rational) refined index has been discovered by Joyce and Song \cite{Joyce:2008pc}, and Kontsevich and Soibelman \cite{Kontsevich:2008fj} in the mathematical literature. It has also been rederived by Manschot, Pioline and Sen by using the localization in the Coulomb branch \cite{Manschot:2010qz}. In this appendix, we give the specific examples of the wall crossing formul{\ae} in order to compare with our results.

Let us take $\gamma=M\gamma_1+N\gamma_2$, where $\gamma_{1}, \gamma_2$  are primitive charge vectors and $M,N$ are positive integers.
The rational refined index in the generic wall crossing formula is defined by
\begin{align}
\bar{\Omega}_{\mathrm{ref}}(\gamma,y)\equiv \sum_{m|\gamma}\frac{y-y^{-1}}{m(y^m-y^{-m})}\Omega_{\mathrm{ref}}(\gamma/m,y^m), \label{eq:rational}
\end{align}
where the sum runs over all positive integers $m$ such that $\gamma/m$ lies in the charge lattice. We denote the change of the rational refined index across the wall by 
\begin{align}
 \Delta \bar{\Omega}_{\mathrm{ref}}(\gamma,y)\equiv \bar{\Omega}^-_{\mathrm{ref}}(\gamma,y)-\bar{\Omega}^+_{\mathrm{ref}}(\gamma,y),
\end{align}
where $\bar{\Omega}^\pm_{\mathrm{ref}}(\gamma,y)$ are rational refined indices
on two sides of the wall  (in chambers $c^\pm$). In the following, we deal with the cases of $(M,N)=(2,2)$ and  $(2,3)$ to compare with our results.

The wall crossing formal{\ae} for $(M,N)=(2,2)$ and  $(2,3)$ are  given by (A.2) and (A.4) of \cite{Manschot:2010qz}, respectively. Using the definition of the rational refined index (\ref{eq:rational}), the wall crossing formal{\ae} can be written by the usual refined indices. Setting $\Omega_{\mathrm{ref}}^+(N\gamma_1,y)=\Omega_{\mathrm{ref}}^+(N\gamma_2,y)=\delta_{N,1}$, the wall crossing formal{\ae} become
\begin{align}
\Delta \bar{\Omega}_{\mathrm{ref}}(2\gamma_1+2\gamma_2,y)&=-\frac{y^{-4k+7}}{2(1+y^2)}\frac{1-y^{4k}}{1-y^4}
\left(\frac{2(1-y^{2k-2})^2}{(1-y^2)^2}-y^{2k-4}\right), \label{eq:22ref} \\
\Delta \bar{\Omega}_{\mathrm{ref}}(2\gamma_1+3\gamma_2,y)&=\frac{y^{-6k+12}}{(1-y^2)(1-y^4)^2(1-y^6)} \notag \\
 &\quad \times \Big[y^{12k-8}-y^{8k-8} (2+3y^2+3y^4+y^6)+y^{6k-8}(1+y^2)^4 \notag \\
 &\qquad -y^{4k-6}(1+3y^2+3y^4+2y^6)+1\Big], \label{eq:23ref}
\end{align}
where we have set $\gamma_{12}=-k$ following the convention of \cite{Manschot:2010qz}.
For $k=1,2,3,4,5$, the formula (\ref{eq:22ref})  reduces to
\begin{align}
 \Delta \bar{\Omega}_{\mathrm{ref}}(2\gamma_1+2\gamma_2,y)|_{k=1}&=\frac{y}{2(1+y^2)}, \label{eq:app221}\\
 \Delta \bar{\Omega}_{\mathrm{ref}}(2\gamma_1+2\gamma_2,y)|_{k=2}&=-\frac{y^{-1}}{2(1+y^2)}(1+y^4), \\
 \Delta \bar{\Omega}_{\mathrm{ref}}(2\gamma_1+2\gamma_2,y)|_{k=3}&=-\frac{y^{-5}}{2(1+y^2)}(1+y^4+y^8)(2+3y^2+2y^4), \\
 \Delta \bar{\Omega}_{\mathrm{ref}}(2\gamma_1+2\gamma_2,y)|_{k=4}&=-\frac{y^{-9}}{2(1+y^2)}(1+y^4+y^8+y^{12})\notag\\
 &\qquad\qquad\times (2+4y^2+5y^4+4y^6+2y^8), \\
 \Delta \bar{\Omega}_{\mathrm{ref}}(2\gamma_1+2\gamma_2,y)|_{k=5}&=-\frac{y^{-13}}{2(1+y^2)}(1+y^4+y^8+y^{12}+y^{16})\notag\\
 &\qquad\qquad\times(2+4y^2+6y^4+7y^6+6y^8+4y^{10}+2y^{12}), \label{eq:app225}
\end{align}
and (\ref{eq:23ref}) becomes
\begin{align}
 \Delta \bar{\Omega}_{\mathrm{ref}}(2\gamma_1+3\gamma_2,y)|_{k=1}&=0, \label{eq:app231}\\
 \Delta \bar{\Omega}_{\mathrm{ref}}(2\gamma_1+3\gamma_2,y)|_{k=2}&=1, \\
 \Delta \bar{\Omega}_{\mathrm{ref}}(2\gamma_1+3\gamma_2,y)|_{k=3}&=\frac{1}{y^6}(1+y^2+3y^4+3y^6+3y^8+y^{10}+y^{12}), \\
 \Delta \bar{\Omega}_{\mathrm{ref}}(2\gamma_1+3\gamma_2,y)|_{k=4}&=\frac{1}{y^{12}}(1+y^2+3y^4+4y^6+7y^8+8y^{10}+10y^{12} \notag \\
 &\qquad\qquad\qquad+8y^{14}+7y^{16}+4y^{18}+3y^{20}+y^{22}+y^{24}), \\
 \Delta \bar{\Omega}_{\mathrm{ref}}(2\gamma_1+3\gamma_2,y)|_{k=5}&=\frac{1}{y^{18}}(1+y^2+3y^4+4y^6+7y^8+9y^{10}\notag\\
 &\qquad+14y^{12}+16y^{14}+20y^{16} +20y^{18}+20y^{20}+16y^{22}\notag \\
 &\qquad\quad+14y^{24}+9y^{26}+7y^{28}+4y^{30}+3y^{32}+y^{34}+y^{36}). \label{eq:app235}
\end{align}

\section{Deformation of FI parameters} \label{mod}

We here explain how to deform the FI parameters. At first, let us consider the general FI parameters as follows:
\begin{align}
\zeta_1\mathbf{1}_2 &\to 
\begin{pmatrix}
\zeta_1^{(1)} & 0 \\
0 & \zeta_1^{(2)}  
\end{pmatrix}, \label{eq: FIgen1} \\
\zeta_2\mathbf{1}_2 &\to
\begin{pmatrix}
\zeta_2^{(1)}  & 0 \\
0 & \zeta_2^{(2)} 
\end{pmatrix}, \label{eq: FIgen2} 
\end{align}
where $\zeta_1^{(1)}+\zeta_1^{(2)}+\zeta_2^{(1)}+\zeta_2^{(2)}=0$ due to the $\theta$-stability condition $\theta (\mathbf{N})=0$. Since the original FI-paramaters are given by $\zeta_1^{(1)}=\zeta_1^{(2)}=\zeta,~\zeta_2^{(1)}=\zeta_2^{(2)}=-\zeta$, we assume 
\begin{align}
 \zeta_1^{(1)}\simeq \zeta_1^{(2)}\simeq \zeta,~~~~~~~~~\zeta_2^{(1)}\simeq \zeta_2^{(2)}\simeq -\zeta.
\end{align}
Now we solve the F-term and D-term constraints on the each fixed point.

\begin{enumerate}
\item[(v1)] The fixed point (\ref{u2u2 BRST})

The D-term constraint is given by
\begin{align}
 \frac{1}{2g^2}|z^1_{12}|^2 + |q^a_{11}|^2 &= \zeta_1^{(1)}, \label{Dtermgen1} \\
-\frac{1}{2g^2}|z_{12}^1|^2 + |q^a_{22}|^2 &= \zeta_1^{(2)}, \label{Dtermgen2}\\
\frac{1}{2g^2}|z^2_{12}|^2 - |q^a_{11}|^2 &= \zeta_2^{(1)}, \label{Dtermgen3}\\
-\frac{1}{2g^2}|z^2_{12}|^2 - |q^a_{22}|^2 &=\zeta_2^{(2)}. \label{Dtermgen4}
\end{align}
One of these equations is not independent because of the $\theta$-stability condition. 
The F-term constraint is
\begin{align}
q_{22}^az^1_{12}=q_{11}^az^2_{12}.
\end{align}
Solving the constraints, we find the solutions are given by
\begin{align}
 |q_{11}^a|^2&=\frac{\zetaa(\zetaa+\zetab)}{\zetaa-\zetad}, \\
 |q_{22}^a|^2&=-\frac{\zetad(\zetaa+\zetab)}{\zetaa-\zetad}, \\
 |z^1_{12}|^2&=2g^2\frac{\zetaa(\zetaa+\zetac)}{\zetaa-\zetad}, \\
 |z^2_{12}|^2&=-2g^2\frac{\zetad(\zetaa+\zetac)}{\zetaa-\zetad}.
 \end{align}
Thus, if $\zetaa+\zetac >0$, these constraints are satisfied and the fixed point (\ref{u2u2 BRST}) contributes to the partition function.

\item[(v2)] The fixed point (\ref{vectorfail1})

The solutions of the constraints are given by
\begin{align}
|q_{11}^a|^2&=-\frac{\zetac(\zetaa+\zetab)}{\zetab-\zetac}, \\
|q_{22}^a|^2&=\frac{\zetab(\zetaa+\zetab)}{\zetab-\zetac}, \\
|z^1_{21}|^2&=-2g^2\frac{\zetab(\zetaa+\zetac)}{\zetab-\zetac}, \\
|z^2_{21}|^2&=2g^2\frac{\zetac(\zetaa+\zetac)}{\zetab-\zetac}.
\end{align}
Thus, if $\zetaa+\zetac<0$, the constraints are satisfied. Therefore, if we choose the fixed point  (\ref{u2u2 BRST}), we can not take the fixed point (\ref{vectorfail1}).

\item[(v3)] The fixed point (\ref{eq:vectoradd})

The solutions of the constraints are given by
\begin{align}
|q_{12}^a|^2&=\frac{\zetaa(\zetaa+\zetab)}{\zetaa-\zetac}, \\
|q_{21}^a|^2&=-\frac{\zetac(\zetaa+\zetab)}{\zetaa-\zetac}, \\
|z^1_{12}|^2&=-2g^2\frac{\zetaa(\zetab+\zetac)}{\zetaa-\zetac}, \\
|z^2_{21}|^2&=2g^2\frac{\zetac(\zetab+\zetac)}{\zetaa-\zetac}.
\end{align}
Thus, if $\zetab+\zetac<0$, the constraints are satisfied.

\item[(v4)] The fixed point (\ref{vectorfail2})

The solutions of the constraints are given by
\begin{align}
|q_{12}^a|^2&=-\frac{\zetad(\zetaa+\zetab)}{\zetab-\zetad}, \\
|q_{21}^a|^2&=\frac{\zetab(\zetaa+\zetab)}{\zetab-\zetad}, \\
|z^1_{21}|^2&=2g^2\frac{\zetab(\zetab+\zetac)}{\zetab-\zetad}, \\
|z^2_{12}|^2&=-2g^2\frac{\zetad(\zetab+\zetac)}{\zetab-\zetad}.
\end{align}
Thus, if $\zetab+\zetac>0$, the constraints are satisfied. Therefore, we can not take both the fixed points (\ref{eq:vectoradd}) and (\ref{vectorfail2}) at the same time.
\end{enumerate}

\begin{enumerate}
\item[(c1)] The fixed point $\Phi^*[^{ab}_c]$

Solving the D-term constraint, we find
\begin{align}
|q_{11}^a|^2&=-(\zetab+\zetac), \\
|q_{12}^b|^2&=-\zetad, \\
|q_{21}^c|^2&=\zetab.
\end{align}
Thus, if $\zetab+\zetac<0$, the constraint is satisfied.

\item[(c2)] The fixed point $\Phi^*[^{ab}_{~c}]$

The solutions of the D-term constraint are
\begin{align}
|q_{11}^a|^2&=-\zetac, \\
|q_{12}^b|^2&=\zetaa+\zetac, \\
|q_{22}^c|^2&=\zetab.
\end{align}
Thus, if $\zetaa+\zetac>0$, the constraint is satisfied.

\item[(c3)] The fixed point $\Phi^*[^{~a}_{bc}]$

The solutions of the D-term constraint are
\begin{align}
|q_{12}^a|^2&=\zetaa, \\
|q_{21}^b|^2&=-\zetac, \\
|q_{22}^c|^2&=\zetab+\zetac.
\end{align}
Thus, if $\zetab+\zetac>0$, the constraint is satisfied. Thus, we can not take both $\Phi^*[^{ab}_c]$ and $\Phi^*[^{~a}_{bc}]$ at the same time.

\item[(c4)] The fixed point $\Phi^*[^{a}_{bc}]$

The solutions of the D-term constraint are
\begin{align}
|q_{11}^a|^2&=\zetaa, \\
|q_{21}^b|^2&=-(\zetaa+\zetac), \\
|q_{22}^c|^2&=-\zetad.
\end{align}
Thus, if $\zetaa+\zetac<0$, the constraint is satisfied. Thus, we can not take both $\Phi^*[^{ab}_{~c}]$ and $\Phi^*[^{a}_{bc}]$ at the same time.
\end{enumerate}

The conditions for the FI parameters and the fixed points which we should choose are summarized in Table \ref{fi}. If we take the FI parameters as
$\zetaa=\zeta+\delta,~ \zetab=\zeta-\delta, ~\zetac=\zetad=-\zeta~(\delta \ll\zeta)$, 
the proper fixed points are (v1), (v3), (c1), and (c2). 
\begin{table}
\centering
\begin{tabular}{l|l|l}
& $\zetaa+\zetac>0$ & $\zetaa+\zetac<0$ \\ \hline 
$\zetab+\zetac>0$ & (v1), (v4), (c2), (c3) & (v2), (v4), (c3), (c4) \\ \hline
$\zetab+\zetac<0$ & (v1), (v3), (c1), (c2) & (v2), (v3), (c1), (c4) 
\end{tabular}
\caption{The conditions for the FI parameters and the proper fixed points.}
\label{fi}
\end{table}

\end{document}